\documentclass[a4paper,11pt,openbib,final,footheight=-5pt]{scrartcl}
\usepackage[a4paper,includehead,top=20mm,bottom=20mm,right=20mm,left=20mm]{geometry}

\makeatletter
\DeclareOldFontCommand{\bf}{\normalfont\bfseries}{\mathbf}
\DeclareOldFontCommand{\it}{\normalfont\itshape}{\mathit}
\makeatother

\usepackage[utf8]{inputenc} 

\usepackage[T1]{fontenc}
\usepackage[american]{babel}

\usepackage{amsmath, amsxtra, amssymb, amsthm}

\usepackage{amsfonts, amssymb, amsbsy}
\usepackage{eucal}
\usepackage{csquotes}

\usepackage{bm, bbm, upgreek}
\usepackage{booktabs, multirow}
\usepackage{graphicx}

\usepackage{subfig}

\DeclareMathAlphabet{\mathscr}{OT1}{pzc}{m}{it}

\usepackage{newtxtext}
\usepackage{lscape}
\usepackage[table]{xcolor}
\usepackage{multirow,tabularx}
\usepackage{setspace}
\usepackage{listings}
\usepackage{booktabs}
\usepackage{indentfirst}

\usepackage{eurosym}

\renewcommand{\t}[1]{\bm{#1}}

\newcommand{\p}{\partial}
\renewcommand{\d}{\, \mathrm d }

\newcommand{\del}{\updelta}
\newcommand{\dd}[2]{\frac{\d #1}{\d #2}}
\newcommand{\pd}[2]{\frac{\p #1}{\p #2}}
\newcommand{\comma}{\ , \quad }
\newcommand{\B}{\mathcal B}
\newcommand{\Ha}{\mathcal H}

\newcommand{\eps}{\varepsilon}

\newcommand{\Dt}[1]{#1^{\scalebox{0.5}{\textbullet} } }
\newcommand{\DDt}[1]{#1^{\scalebox{0.5}{\textbullet}\scalebox{0.5}{\textbullet} } }

\newcommand{\DDDDt}[1]{#1^{\scalebox{0.5}{\textbullet}\scalebox{0.5}{\textbullet}\scalebox{0.5}{\textbullet}\scalebox{0.5}{\textbullet} } }

\newcommand{\RRef}{\text{\tiny{ref.}}}

\newcommand{\begeq}{\begin{equation}\begin{gathered}}
\newcommand{\eqend}{\end{gathered}\end{equation}}
\newcommand{\begal}{\begin{equation}\begin{aligned}}
\newcommand{\alend}{\end{aligned}\end{equation}}

\renewcommand{\L}{\mathcal L}

\newcommand{\Om}{\Omega}

\newcommand{\J}{\mathcal I}
\newcommand{\Curr}{\mathcal J}

\newcommand{\PP}{\mathscr P}
\newcommand{\RR}{\mathscr R}

\title{\huge Energy based methods applied in mechanics by using the extended Noether's formalism}
\author{Bilen Emek Abali$^{1}$\thanks{Corresponding author, ORCID: 0000-0002-8735-6071, email: bilenemek@abali.org} 
\\[0.05in]
\small $^1$Uppsala University, \\[-0.1in]
\small Division of Applied Mechanics, \\[-0.1in]
\small Department of Materials Science and Engineering, \\[-0.1in]
\small Box 534, SE-751 21 Uppsala, Sweden
}

\begin{document}
\maketitle

\begin{abstract}
Physical systems are modeled by field equations; these are coupled, partial differential equations in space and time. Field equations are often given by balance equations and constitutive equations, where the former are axiomatically given and the latter are thermodynamically derived. This approach is useful in thermomechanics and electromagnetism, yet challenges arise once we apply it in damage mechanics for generalized continua. For deriving governing equations, an alternative method is based on a variational framework known as the extended \textsc{Noether}'s formalism. Its formal introduction relies on mathematical concepts limiting its use in applied mechanics as a field theory. In this work, we demonstrate the power of extended \textsc{Noether}'s formalism by using tensor algebra and usual continuum mechanics nomenclature. We demonstrate derivation of field equations in damage mechanics for generalized continua, specifically in the case of strain gradient elasticity.
\end{abstract}


\section{Introduction}

In rational continuum mechanics, we axiomatically start with the balance equations in a formal manner established in \cite{truesdell_toupin}. For thermomechanics there are balance equations of mass, momenta, and energy. For electromagnetism there are \textsc{Faraday} relation (balance of magnetic flux) and balance of electric charge, they lead to \textsc{Maxwell}'s equations. These balance equations are called ``universal'' since they hold for all materials. For a specific application with known materials, equations for stress, heat flux, (internal or free) energy, electric current, electromagnetic force, charge and current potentials are yet to be defined by constitutive equations in order to close the system of equations. The constitutive equations are needed for establishing the material specific behavior into the system. 

Starting with \cite{eckart1940I, eckart1940II, eckart1940III}, the thermodynamics of irreversible processes has been used for obtaining the constitutive equations in a formal way. There are ample methods in the literature, we may name at least four famous approaches: The \textsc{Coleman--Noll} procedure \cite{ColemanNoll1963}, \textsc{M\"uller}'s rational thermodynamics \cite{mueller1973}, non-equilibrium thermodynamics \cite{groot1984}, and extended thermodynamics \cite{muller1993extended}. Thermodynamical approaches consider bulk properties; for surface phenomena such as crack propagation or surface polarization, additional assumptions or models need to be suggested. In the case of surfaces, additional axiomatic balance equations may be used for singular surfaces \cite{muller2023electrodynamics}. When it comes to edge effects, further complications arise and there is simply not one method covering all systems in  generalized continua. In short, for applications involving first space derivatives (gradients) of unknowns---the method is useful \cite{014}, where we start directly with balance equations and derive all the rest. If an application demands higher gradients to cover surface and edge effects (and even beyond) then we need a generalization of this method, technical challenges arise. 

The formal difficulties aside, definition of internal or free energy is challenging as well. Energy is a directly measurable quantity and a theory based on energy is known as the variational formulation with roots over many centuries long \cite{dell2014complete,dell2018complete}. In continuum mechanics, variational formulation has been suggested in different settings, in hydrodynamics \cite{seliger1968variational,auffray2015analytical}, fluid-structure interactions \cite{kock1991fluid,placidi2008variational}, multiphysics \cite{prix2004variational,bekenstein2000conservation}, dynamics \cite{altenbach2014vibration,eremeyev2019dynamic}, mechanics with dissipation \cite{placidi2015variational,ciallella2021rate}, and for discrete structures \cite{steigmann1996variational}. Specifically for crack propagation, variational approaches exist in quasi-static cases \cite{bourdin2008variational}. This approach uses an energy concept and may be extended to generalized mechanics, yet it causes technical difficulties in the case of applied surface or volume forces \cite{francfort1998revisiting}. In classical mechanics, energy integral in time is called action. Invariance of action is postulated by using a \textsc{Lagrange} function. Although the method is useful, its motivation is formal \cite{vujanovic1975variational}. This \textsc{Lagrange} function leads to \textsc{Euler--Lagrange} equations that are the field equations for the bulk. In the case of bulk quantities, classical mechanics \cite{lanczos1949variational,gelfand1963calculus,goldstein2002classical} is utilized for displacement in ``regular'' systems, i.e. with no singularities. 

Topological methods by using geometric formulations \cite{arnold2021topological,marsden2001variational} have been exploited for developing a geometric continuum mechanics \cite{morrison1984bracket, grmela1986bracket, holm1998euler, gay2018lagrangian}, which are applied in multiphysics as well as in multiscale theories \cite{beris1994thermodynamics,pavelka2018multiscale}. Such formulations are developed for regular domains, in simpler terms, the underlying mathematical structure needs smooth functions that is only possible if the continuum body has no voids or cracks. A crack may be seen as a singular surface in the sense that the displacement value on both sides of the surface is different. Thus, there is a jump in the displacement field across this ``singular'' surface. We call this singularity a ``defect'' or ``fracture'' in the material. 

In elasticity, a propagating crack is modeled by forces that are acting on this defect \cite{eshelby1951force}. These so-called material (or configurational) forces are conceptually beneficial for generating a so-called material stress and an energy-momentum tensor as introduced in \cite{eshelby1975elastic}. This method provides a model for the surface related phenomena. The energy-momentum tensor is written by using a space-time continuum. This approach has been proven to be adequate in the special and then general relativity \cite{dirac}, where a spatial space is used in the formulation. A spatial space is coordinates pointing at locations with or without massive particles occupying this locations. This concept is well-known in fluid dynamics with a control volume, also called a \textsc{Euler}ian approach, where space coordinates are fixed in the physical space (where we live). We call it a laboratory frame. Motion of coordinates are not related to the motion of massive particles. Often we visualize that the laboratory frame is fixed (not moving). In continuum mechanics, we use a material frame where the space coordinates point to the same massive particle (material point or matter). Motion of coordinates means the motion of massive particles. As the matter may move, a continuum body is deforming, we may think of a co-moving frame for space. The four-dimensional notation has been used in the \textsc{Eshelby} energy-momentum tensor by means of a material frame. In non-relativistic approaches, we skip a distinction between time defined on a laboratory or co-moving frame. By decomposing the energy-momentum tensor into a time and space part, the so-called \textsc{Eshelby} stress tensor is obtained and used in connection with crack modeling in different formulations \cite{rice1968path, altenbach1990konzepte, kroner1993configurational, maugin1995material, gurtin1996configurational} and also in applications with numerical examples \cite{steinmann2000application, kienzler2002fracture, li2006dual, kuhn2010continuum, agiasofitou2017micromechanics, perez2017fracture}. For the history of \textsc{Eshelby} stress tensor and its use in variational formulation, we refer to \cite{maugin2016fracture}. For technical details in surface phenomena \cite{dell1987derivation} and \textsc{Eshelby} stress tensor's relation to fracture mechanics, we refer to \cite{kienzler2012mechanics}. 

Displacement is the primitive function that we calculate in mechanics. In continuum mechanics, functions are regular without singularities. Crack is indeed causing a jump of the displacement, since the bond between material particles has been lost, and its thickness is at a smaller length-scale such that we consider this singularity as a fictitious surface (without its own mass). The \textsc{Eshelby} stress tensor is used in \cite{buratti2003} in a systematic study in order to define an entropy production on the singular surface. This excellent idea leads naturally to constitutive equations for the velocity of the singular surface that is tantamount to the crack in physics; in the literature, there is no such consideration for suggesting a relation and its experimental verification. Another novel approach is presented in \cite{wolff2018application}, where \textsc{M\"uller}'s rational thermodynamics is used to obtain a balance equation with the necessary constitutive equation for the damage parameter. A differential equation related to the \textsc{Eshelby} stress tensor is without doubt a possible modeling tool for crack propagation  \cite{mueller2002material,ruter2006goal}. 

In a first-order continuum, in the case of elasticity, we know the governing equations including \textsc{Eshelby} stress tensor. When it comes to a generalization, for example higher-order theories or in multiphysics, it becomes challenging to motivate if and how \textsc{Eshelby} stress tensor needs to be modified. Therefore, we follow another approach and derive field equations by using \textsc{Noether}'s formalism in continuum mechanics. In this approach, all governing equations, including \textsc{Eshelby} stress tensor, are acquired by applying \textsc{Noether}'s formalism. Herein we exploit this approach by using a continuum mechanics jargon known as \textsc{Ricci} calculus. The manuscript comprises a detailed motivation of the formalism: 
\begin{itemize}
\item
In Sect.~\ref{sect:variation}, we discuss about variation of fields, which is the core of the subject and is discussed in an abstract setting. We give examples to well-known transformation rules for clarifying their role in this formulation. For the sake of consistency, we include possibly well known concepts from variational calculus and tensor algebra. 
\item
In Sect.~\ref{sect:noether}, we explain the extended \textsc{Noether}'s theorem for fields. We emphasize that we demonstrate (discrete) systems as in dynamics depending on time; and also (continuous) systems in space and time. It is of utmost importance to distinguish that many variational formulations utilize a spatial space. Historically, \textsc{Noether}'s theorem has been applied for such systems mainly in electromagnetism, since there, the fields propagate with or without matter. But herein, we focus on a continuum body, where space is defined as attached to and co-moving with the matter. Therefore, we may call it a "configurational" space, see \cite{singh2021pseudomomentum} for the history of this distinction. We obtain the so-called \textsc{Rund--Trautman} identity in this configurational space.
\item
In Sect.~\ref{sect:conservation}, we obtain the usual \textsc{Euler--Lagrange} equations. We provide the derivation of \textsc{Noether}'s current leading to conservation (balance) laws, by using \textsc{Euler--Lagrange} equations and \textsc{Rund--Trautman} identity. The formalism is utilized in an abstract manner in order to employ for generalized continuum.
\end{itemize}
An important outcome of this work is the clear differing in governing equations and their derivations. In this way, we enable a more broader use of this formalism. The \textsc{Noether}'s formalism leads to \textsc{Euler--Lagrange} equations. This approach is often used in classical mechanics for discrete systems, but we use it herein for obtaining field equations. The balance laws are acquired with this formalism. Usually, after acquiring the balance laws, the \textsc{Noether}'s formalism is taken aside. We do the contrary and continue with the extended \textsc{Noether}'s formalism that leads to \textsc{Rund--Trautman} identity. This identity contains not only balance laws yet also \textsc{Eshelby} stress tensor, where the latter leads to a so-called $J$-integral used in damage mechanics. Therefore, we exploit the extended \textsc{Noether}'s formalism in order to obtain field equations in damage mechanics, first, for classical elasticity, then, for strain gradient elasticity:
\begin{itemize}
\item
In Sect.~\ref{sect:first.order}, we give examples of well-known elastodynamics by using the extended \textsc{Noether}'s formalism. In this way, it is clarified that the balance of momentum is obtained from the \textsc{Euler--Lagrange} equations, balance of energy is acquired from the \textsc{Noether}'s current, and the \textsc{Eshelby} stress tensor is attained by the \textsc{Rund--Trautman} identity.
\item
In Sect.~\ref{sect:second.order}, we apply the same approach for generalized continua. In addition to balance laws for generalized elasticity, we acquire \textsc{Eshelby} stress tensor (and $J$-integral) in the case of strain gradient elasticity.
\end{itemize}

\section{Variation of fields}\label{sect:variation}

We use standard continuum mechanics notation and understand a summation over all repeated indices called \textsc{Einstein} summation convention. Consider a body composed of massive particles. Particle is defined as the smallest observable piece of material consisting of an infinitesimal volume element. This volume element is, however, large enough to neglect any quantum phenomena. The body's image is the continuum domain in the space $R^3$, with (contravariant) coordinates $x^i$, where $i=1,2,3$. Coordinates label the particles of the body; in other words, we use the particle coordinates analogous as in \cite{eckart1960}. We use time, $t$, and space, $(x^1,x^2,x^3)$; the numerical values of coordinates do not change as they denote the same massive particle. As the numerical values of $x^i$ do not change for the same particle, the particle rests in this ``material'' frame. Since the equilibrium is defined by referring the material frame, we call it the \textit{reference} frame. It may be called the \textit{inertial} frame with respect to which the dynamics of the system is prescribed. An inertial frame is basically the coordinate system labeling particles, since the definition of the inertial frame is the one where the mass rests,\footnote{From a philosophical point of view, the co-moving frame shall not be called an inertial frame since we cannot distinguish between the forces due to the gravitational forces and due to the acceleration (inertial).} and thus, does not carry any inertial forces \cite{disalle2009}. 

Motion is modeled by means of two distinct processes. First, the particle may deviate from its equilibrium position, this displacement is fully recoverable. The best example is in elasticity. An elastic deformation is such that the particles change their positions by deviating from the reference frame (equilibrium); after unloading, the particles turn to the same equilibrium position. Reference frame itself---equivalently the equilibrium position of the continuum body---may be altered irreversibly. If the motion of the body is such that the system obtains another equilibrium position upon unloading, as in plasticity, then we need different field equations defining the deviation from the equilibrium and the motion (evolution) of the reference frame. 

We aim for a general formulation that employs transformation properties of tensors. We start with the continuous transformation that forms the basis of the \textsc{Hamilton--Jacobi} equation in the \textsc{Noether}ian formalism. This variational principle leads to governing equations as introduced formally by \cite{noether1} and \cite{noether2} we refer to \cite{kosmann2011noether} for the historical remarks about \textsc{Noether}'s theorem. The variational principle is based on the variational calculus that is the algebra of functionals and their extremal values. Suppose that we define a (tensor of rank $0$) function $L(t;x^i,\p x^i/ \p t)$ depending on an independent variable, $t$, and on variables depending on this independent variable, $x^i=x^i(t)$ and $\p x^i / \p t = \frac{\p x^i}{\p t} (t)$. Its integral over the independent variable is called action: 
\begeq
\J = \int L \d t \comma
\eqend
which becomes a form by rewriting,
\begeq
\J = \int \dd{U}{t}\d t = \int \d U \ .
\eqend
This so-called first integral in $U$ means that the path of the integral does not matter, transformation in $t$ leaves its value unchanged. According to the theory of invariants the differential form $\d U$ is an invariant: Its transformation in $x^i$ fails to change its numerical value. In short, action is invariant under transformation of $t$ and $x^i$. We discuss this formal property going back to \cite{collins1899} in more detail in what follows. 

An important side note is the invariants and their use in mechanics. For example in hyperelasticity, deformation energy is modeled by invariants (of strain) since they do not change under coordinate transformations. The values of the (strain) tensor change by following the transformation rules. Although the components of a tensor change, the invariants remain the same. Therefore, by using the invariants, we obtain an energy definition independent of coordinate systems.

The transformation is arbitrary\footnote{Often, its study is conducted by using differential forms \cite{flanders}. We will not make much use of this so called exterior calculus and use the fact  that tensors in oblique coordinate systems under (affine) transformations produce same calculus as the invariant theory of (differential) forms  \cite[\textsection 9]{pauli2}.} and $L$ is called \textsc{Lagrange}an. In order to understand, what the transformation means, we briefly introduce a continuous transformation. We use a function taking the variable $x^i$ and transforms along $\eps$, as follows:
\begeq\label{transformation.general}
x^{i'}=\zeta^i(\t x,\eps) \ .
\eqend
We provide some examples in Appendix \ref{appendix}. The general framework is used for an arbitrary but linear transformation. When we expand $\zeta$ in \textsc{Taylor} series, around $\eps=0$, then a first order approximation for the transformation reads
\begeq\label{first.order.approximation}
x^{i'}= x^i + \eps \pd{\zeta^i}{\eps}\Big|_{\eps=0} + \mathcal O(\eps^2) \approx x^i+\eps\xi^i(\t x) \comma
\eqend
where the coefficients to the first power, $\xi^i$, are called the \textit{generators} of the transformation \cite[\textsection~4.1]{neuenschwander2010}. As at the limit $\eps\rightarrow 0$, the approximation becomes accurate, $\eps$ is often introduced as a ``small'' parameter. If we define a tensor of rank one and weight zero, $\t A$, in the space $R^3$, then the variation of the tensor's (contravariant) components becomes, due to the change of (contravariant) coordinates,
\begeq
x^{i'}=x^i+\eps\xi^i \comma \\
A^i\Big|_{x^{j'}} =A^i\Big|_{x^j} + \del A^i\Big|_{x^j} = \pd{x^{i'}}{x^j} A^j\Big|_{x^j} = \Big( \delta^i_j + \eps \pd{\xi^i}{x^j} \Big) A^j\Big|_{x^j} \comma  \\
A^i + \del A^i = \Big( \delta^i_j + \eps \pd{\xi^i}{x^j} \Big) A^j \comma \del A^i= \eps \pd{\xi^i}{x^j} A^j \ .
\eqend
In other words, the variation of $A^i$ means 
\begeq
\del A^i = A^i\Big|_{x^{j'}} - A^i\Big|_{x^j} = A^i(x^{j'}) - A^i(x^j) \ .
\eqend
Standard tensor calculus rules apply; the transformation is invertible,
\begeq
\pd{x^{i'}}{x^j} \pd{x^j}{x^{k'}}=\delta^{i'}_{k'}=\delta^i_k \ .
\eqend
We find the variation of covariant components of the same tensor of rank one \cite[\textsection~23]{pauli2} in a straight-forward manner
\begeq
x^i=x^{i'}-\eps \xi^i \comma \\
A_i\Big|_{x^{j'}}=A_i\Big|_{x^j} + \del A_i\Big|_{x^j}=A_j \pd{x^j}{x^{i'}}=A_j\Big( \delta^j_i - \eps\pd{\xi^j}{x^{i'}} \Big) \comma \del A_i = -\eps A_j \pd{\xi^j}{x^i} \ .
\eqend
Analogously, for a tensor of rank two and weight zero, we obtain
\begeq
\del A^{ik}=\eps \pd{\xi^i}{x^j} A^{jk} + \eps \pd{\xi^k}{x^j}A^{ij} \comma \\
\del A^i_{\ k}=\eps \pd{\xi^i}{x^j} A^j_{\ k} - \eps \pd{\xi^j}{x^k}A^i_{\ j} \comma \\
\del A_{ik}=-\eps \pd{\xi^j}{x^i} A_{jk} - \eps \pd{\xi^j}{x^k}A_{ij} \ .
\eqend
As we have determined the transformation of co- and contravariant components of tensors, we start applying the formalism regarding transformation of \textsc{Lagrange}an. 
\section{Extended \textsc{Noether}'s theorem}\label{sect:noether}
Suppose that $L(t;\chi^i,\p \chi^i/\p t)$ is defined by an independent variable, $t\in [a,b]$, by variables, $\chi^i=\chi^i(t)$, and their derivatives, $\p \chi^i/\p t=\pd{\chi^i}{t} (t)$, with respect to the independent variable. We define the transformation, $t'=t+\eps\tau$ and $\chi^{i'}=\chi^i+\eps \xi^i$, with the corresponding generators, $\tau, \xi^i$. If $L$ is a scalar, its numerical value does not change---it is invariant---with respect to a particular choice of an infinitesimal transformation with the generators, $\tau$, $\xi^i$, as follows:
\begeq
L(t;\chi^i,\p \chi^i/\p t)=L(t';\chi^{i'},\p \chi^{i'}/\p t') \ .
\eqend
We use a simplified notation $L'=L(t';\chi^{i'},\p \chi^{i'}/\p t')$ and rewrite $L=L'$. The (action) functional, $\J=\int L\d t$, has its variation
\begeq
\del \J = \J'-\J = 
\int L\Big(t';\chi^{i'},\pd{\chi^{i'}}{t'} \Big) \d t' 
- \int L \Big(t;\chi^i,\pd{\chi^i}{t}\Big) \d t 
= \\
= \int \bigg( L\Big(t';\chi^{i'},\pd{\chi^{i'}}{t'} \Big) \frac{\d t'}{\d t}- L \Big(t;\chi^i,\pd{\chi^i}{t}\Big) \bigg)\d t  \ .
\eqend
As $L=L'$ or in other words, $\del L=0$, we expect that $\del \J$ is independent of $t$, $\chi^i$. The trivial way is to choose, $\del\J=0$. As the theory is at first order, i.e. in $\eps^1$, owing to Eq.\,\eqref{first.order.approximation}, we acquire $\del\J \rightarrow 0$ faster than $\eps\rightarrow 0$. By choosing a small $\eps$, linear in $\eps$ is an accurate theory in first order. Therefore, we may have a linear in $\eps$ difference,
\begeq
\del \J=\int \eps \d F=\int \eps \frac{\d F}{\d t} \d t \ ,
\eqend
where $F$ is an arbitrary (but given) function. The latter brings an important equation:
\begeq
L\Big(t';\chi^{i'},\pd{\chi^{i'}}{t'} \Big) \frac{\d t'}{\d t}- L \Big(t;\chi^i,\pd{\chi^i}{t}\Big) = \eps \frac{\d F}{\d t} \ .
\label{div.inv.single}
\eqend
This abstract transformation that produces a source term has been used extensively in \cite{helmholtz1887,helmholtz1887ueber} to derive the \textit{principle of least action}. A more general form for the right-hand side fails to exist, since the transformation is linear in $\eps$. Although not written in this way, the formalism goes back to \textsc{Maupertius} and has been used for solving differential equations. This formalism is often called a \textit{canonical transformation}.\footnote{In solving partial differential equations canonical transformation has another meaning of bringing the set of equations into a \textsc{Jordan} normal form. Here the same name is used for a different formalism.} We transform the functional in the parameter $t$ that produces a right-hand side, if $L$ is in unit of energy then the induced $F$ is in momentum times position. 

Now, we generalize the procedure by defining $(t,x^i)$ as independent variables and $\phi^k=\phi^k(t,x^i)$ as variables depending on the independent variables. A \textsc{Lagrange}an density (per space-time) depends on them and their derivatives,
\begeq
\L=\L\Big( t,x^i;\phi^k,\pd{\phi^k}{t},\pd{\phi^k}{x^i}\Big) \comma \J= \int \L \d \Sigma \ ,
\label{L.fields}
\eqend
where $\L$ is in unit of energy per volume. We use an infinitesimal space-time element, $\d\Sigma = \sqrt{g} \d x \d t$, in a frame with the metric tensor, $g_{\mu\nu}$, and its determinant, $g=\det(g_{\mu\nu})$. As aforementioned, this frame may be chosen as laboratory (fixed) or reference (co-moving) frame. Herein, we choose it as the reference frame (material system) and use a standard mapping to an arbitrary frame by means of a transformation. In continuum mechanics, such a mapping is introduced from reference to current frame. Herein, we use a general formulation by using an arbitrary (linear) transformation. We define the reference frame in a Cartesian system, i.e. $\d x\equiv \d x^1 \d x^2 \d x^3$, and thus, the transformed frame is oblique but not curvilinear, $\d x' = \eps^{ijk} \d x^1_{i'}\d x^2_{j'}\d x^3_{k'}$, by using the \textsc{Levi-Civita} symbol, $\eps^{ijk}$. By constructing a four-dimensional space for space-time with its metric (for continuous transformation), $g_{\mu \nu}$, where $\mu,\nu\in \{1,2,3,4\}$, we obtain
\begal\label{cont.trafo}
X^{\mu'}_{\mu} = & \delta^{\mu'}_\mu + \eps\pd{(\tau,\xi^1,\xi^2,\xi^3)^{\mu'}}{(t,x^1,x^2,x^3)^\mu}=
\begin{pmatrix} 
1+\eps\pd{\tau}{t} & \eps\pd{\tau}{x^1} & \eps\pd{\tau}{x^2} & \eps\pd{\tau}{x^3}  \\
\eps\pd{\xi^1}{t} & 1+\eps\pd{\xi^1}{x^1} & \eps\pd{\xi^1}{x^2} & \eps\pd{\xi^1}{x^3} \\
\eps\pd{\xi^2}{t} & \eps\pd{\xi^2}{x^1} & 1+\eps\pd{\xi^2}{x^2} & \eps\pd{\xi^2}{x^3} \\
\eps\pd{\xi^3}{t} & \eps\pd{\xi^3}{x^1} & \eps\pd{\xi^3}{x^2} & 1+\eps\pd{\xi^3}{x^3}
\end{pmatrix}  \comma
\\
J = & \det (X^{\mu'}_\mu)= 1 + \eps \pd{\tau}{t} + \eps \pd{\xi^i}{x^i} + \mathcal O(\eps^2) \comma
\alend
and therefore $\d\Sigma' = J\d\Sigma$ or equivalently $\d x' \d t'=J\d x \d t$. Moreover, we have
\begeq
g_{\mu \nu}=X^{\mu'}_\mu X^{\nu'}_\nu g_{\mu'\nu'} \comma 
g'=\det (g_{\mu'\nu'}) \comma
g =\det (g_{\mu\nu}) \ , 
\eqend
by taking the determinant of the first equation, we obtain
\begal
g = & J J g' \ , \\
\sqrt{g} = & J \sqrt{g'} \ .
\alend
Since the continuous transformations are written as a set with an identity element, they have unique inverse transformations; therefore, they form a \textsc{Lie} group according to the group theory. Of course, we want to have mathematical relations transforming as the coordinate system transforms. Therefore, we use tensors in formulating such relations. A tensor of rank 0 is a scalar. We begin with an assertion that the \textsc{Lagrange} density is such a proper scalar, $\L=\L'$, and observe
\begeq
\J = \int \L \sqrt{g} \d x\d t = \int \L J \sqrt{g'} \d x\d t = \int \L \sqrt{g'} \d x' \d t' = \int \L' \sqrt{g'} \d x' \d t'  = \J' \ .
\eqend
Equivalently, we may begin with a scalar, $\J=\J'$, for a given transformation, and obtain that the \textsc{Lagrange}an is an invariant. This invariance is exploited for the principle of least action as aforementioned, and this formalism is called \textsc{Noether}'s theorem. We continue by building on this formalism and demonstrate an extension of this formalism. 

Let us choose a specific type of generators for the arbitrary transformation, say, \textsc{Lorentz} transformation; $\J$ is called a \textsc{Lorentz} scalar. We may give an analogy to continuum mechanics by neglecting shortly the time integral; consider that $\J=\J'$ means that the energy is the same in both frames. This fact makes sense, but how do we know that the energy density (per volume) is also the same in both frames such that $\L=\L'$ is also fulfilled? Indeed, we may want that the energy density is the same for corresponding material particles, but how do we enforce this case? Now by using this analogy, we ask to answer the question: What if we do have a scalar, $\J'=\J$; however, not a proper scalar, $\L\neq \L'$, by definition, what are the additional conditions to be fulfilled in order to generate a proper scalar, $\L = \L'$ \,?

For a formulation with field equations, we may incorporate the source term in Eq.\,\eqref{div.inv.single} (right-hand side) in two steps. First, we build up the procedure for fields with the \textsc{Lagrange}an in Eq.\,\eqref{L.fields} for arbitrary transformations in space and time. All the transformation is achieved with different generators but one-parameter $\eps$. The variation reads in domain $\Om$ for fields
\begeq
\del \J= \int_{\Om'} \L'\d \Sigma' - \int_\Om \L \d \Sigma = \int_\Om \bigg( \L' J - \L \bigg) \d \Sigma \comma
\eqend
since $ \d\Sigma'=J\d\Sigma $. Second, from \cite{laue1911} we know that a transformation may produce a non-conservative force. In order to implement this concept, we need a clear distinction between conservative and non-conservative forces. A conservative force is derivable from a proper scalar, $S$, which is an invariant, as follows:
\begeq
F^\text{cons.}_i=\pd{S}{x^i} \ .
\eqend
A non-conservative force does not have this special property; thus, in the general case, we may generate a tensor rank 1 from a tensor rank 2, $S_i^k$, by using its derivative
\begeq
F^\text{non-cons.}_i=\pd{S_i^k}{x^k} \ .
\eqend
This property is a general identity, because in tensor calculus, we may always reduce the rank by a derivative. 

In a transformation from one frame (with prime) onto another frame (without prime), a (non-conservative) force may be generated. This fundamental property may be  physically understood as a (virtual) work in case of such a transformation. If we suppose that this transformation is between frames like the current and reference one, the deviation is a virtual displacement: $\del u^i=x^{i'}-x^i$. For this transformation, we need to supply an energy, i.e. a virtual work into the system, $\del A=F_i\del u^i$, where this non-conservative force is measurable on the reference frame. The work is virtual since the transformation between the frames is nothing physical. The choice of frame, where the fields are described and equations are evaluated, has nothing to do with the system itself. Thus, the work caused by the transformation is virtual. However, the force is real and measurable. The direction of transformation from the reference to current gives a minus sign
\begeq
(t,x^i) \rightarrow (t,x^{i'}) \comma \\
x^{i'} = x^i + \del u^i \comma 
J=1+\pd{\del u^i}{x^i} \ , \\
\L=\L' J + \del A  \Rightarrow \del \J = \int_\Omega \big( \L' J - \L \big) \d \Sigma = -\int \del A \d \Sigma \ .
\eqend
This notion is used in elasticity. But the shown principle is applicable for a transformation between laboratory and reference frame as well. Then a transformation from the fixed laboratory frame (control volume) to the co-moving reference frame reads as a positive virtual work on the right-hand side. We are going to use $\pm$ in front of this term, in order to emphasize that both choices are adequate depending on the chosen frame to transform to. This right-hand side is introduced for the first time in \cite{bessel1921} as $\del\J=\eps\pd{F^\mu}{x^\mu}$ that is a simplification and is called a \textit{divergence invariance}. We stick to the more general form with the virtual work. We emphasize that this virtual work is given by a non-conservative force such that it is possible to begin with $\del \J = \pm\del \mathcal R$, where $\del \mathcal R$ is a non-conservative (dissipative) work, also called \textsc{Rayleigh} dissipation. Indeed, the names work, force, and displacement are in harmony with mechanics, but the relation holds true in multiphysics, as also known from dynamics in discrete systems, and they may be called ``generalized'' force or work-conjugate term. In thermodynamics, one often calls them ``thermodynamical'' forces and fluxes.

The invariance of the \textsc{Lagrange}an leads to the \textsc{Rund--Trautman} identity \cite[\textsection~6.5]{neuenschwander2010}. The \textsc{Lagrange}an density $\L'=\L\big(t',x^{i'};\phi^{k'},\pd{\phi^{k'}}{t'},\pd{\phi^{k'}}{x^{i'}}\big)$ depends on the independent variables $(t,x^i)$ and primitive variables $(\phi^k, \pd{\phi^k}{t}, \pd{\phi^k}{x^i})$, which depend on the independent variables. We axiomatically assume that primitive variables exist. We stress that the frame is oblique, hence, we skip distinguishing between covariant and partial space derivatives. The same holds for the time derivative since we measure this in the reference frame. The linear transformations read
\begeq
t'= t + \eps \tau \comma x^{i'}=x^i+\eps \xi^i \comma \phi^{k'}=\phi^k+\eps \varphi^k \ ,
\label{trafo.arbitrary}
\eqend
where all of them are along the same parameter $\eps$. Since the invariance property $L=L'$ asserts the condition to satisfy, $\L' J - \L = \mp \del A$, we can set the variation along one-parameter $\eps$ vanish such as:
\begeq
\dd{}{\eps} \Big( \L' J - \L \pm \del A \Big) \bigg|_{\eps=0} = 0 \ ,
\label{variation.assertion}
\eqend
where we utilize a directed differentiation often called a \textsc{Gateux} derivative. Although we have introduced the virtual work as a physical quantity, $\del A=F_i \del u^i$, with an analogy in mechanics, in the general case, each primitive variable causes virtual work, $\del A=F_i\eps \varphi^i$. The linear transformations depend on $\eps$, and thus, $\L'$ as well as $J$ depend on $\eps$. However, the term $\L$ does not, therefore, we obtain
\begeq
\bigg( \dd{\L'}{\eps} J \bigg)\bigg|_{\eps=0} + \bigg(\L' \dd{J}{\eps}\bigg)\bigg|_{\eps=0} =\mp F_k \varphi^k \ .
\eqend
By using Eq.\,\eqref{cont.trafo}, we obtain
\begeq
\L'\big|_{\eps=0}=\L \ , J=1+\eps\pd{\tau}{t} +\eps\pd{\xi^i}{x^i} \Rightarrow J\Big|_{\eps=0}=1 \ , \dd{J}{\eps}\bigg|_{\eps=0} = \pd{\tau}{t} +\pd{\xi^i}{x^i} \ ,
\eqend
and thus, 
\begeq
\dd{\L'}{\eps} \bigg|_{\eps=0}  + \L \bigg(  \pd{\tau}{t} +\pd{\xi^i}{x^i} \bigg) = \mp F_k \varphi^k \ .
\label{inv.Lag}
\eqend
By using a short notation $\Dt{()}= \p()/\p t$ and $()_{,i}=\p()/\p x^i$, we calculate
\begal
\pd{\phi^{k'}}{t'} \bigg|_{\eps=0} 
=& \pd{ (\phi^k+\eps\varphi^k) }{t} \pd{t}{t'} \bigg|_{\eps=0} 
= \pd{ (\phi^k+\eps\varphi^k) }{t} \pd{(t'-\eps\tau)}{t'} \bigg|_{\eps=0} 
\\
=& \bigg( \pd{\phi^k}{t} + \eps\pd{\varphi^k}{t} \bigg) \bigg( 1 - \eps\pd{\tau}{(t + \eps \tau)}\bigg) \bigg|_{\eps=0} 
=\pd{\phi^k}{t} = \Dt{(\phi^k)} \ ,
\alend
where $t'=t+\eps\tau$ has been explicitly inserted, and obtain
\begal
\dd{}{\eps} \pd{\phi^{k'}}{t'} \bigg|_{\eps=0} 
=& \dd{}{\eps} \Bigg( \bigg( \pd{\phi^k}{t} + \eps\pd{\varphi^k}{t} \bigg) \bigg( 1 - \eps\pd{\tau}{(t + \eps \tau)}\bigg) \Bigg) \Bigg|_{\eps=0} 
=-\pd{\phi^k}{t} \pd{\tau}{t} + \pd{\varphi^k}{t} = \Dt{(\varphi^k)} - \Dt{(\phi^k)} \Dt \tau  \ .
\alend
Analogously, we acquire
\begal
\pd{\phi^{k'}}{x^{i'}} \bigg|_{\eps=0} 
=& \pd{ (\phi^k+\eps\varphi^k) }{x^j} \pd{x^j}{x^{i'}} \bigg|_{\eps=0} 
= \pd{ (\phi^k+\eps\varphi^k) }{x^j} \pd{(x^{j'}-\eps \xi^j)}{x^{i'}} \bigg|_{\eps=0} 
\\
=& \bigg( \pd{\phi^k}{x^j} + \eps\pd{\varphi^k}{x^j} \bigg) \bigg( \delta^j_i - \eps\pd{\xi^j}{(x^i + \eps \xi^i)}\bigg) \bigg|_{\eps=0} 
=\pd{\phi^k}{x^i} = \phi^k_{,i} \ ,
\alend
and
\begal
\dd{}{\eps} \pd{\phi^{k'}}{x^{i'}}  \bigg|_{\eps=0} 
=& \dd{}{\eps} \Bigg( \bigg( \pd{\phi^k}{x^j} + \eps\pd{\varphi^k}{x^j} \bigg) \bigg( \delta^j_i - \eps\pd{\xi^j}{(x^i + \eps \xi^i)}\bigg) \Bigg) \Bigg|_{\eps=0} 
=-\pd{\phi^k}{x^j} \pd{\xi^j}{x^i} + \pd{\varphi^k}{x^i} = \varphi^k_{,i} - \phi^k_{,j} \xi^j_{,i}  \ .
\alend
The first term in Eq.\,\eqref{inv.Lag} reads
\begal\label{variational.part.formulation.term.one}
\dd{\L'}{\eps} \bigg|_{\eps=0} 
=& \dd{\L\big(t',x^{i'};\phi^{k'},\pd{\phi^{k'}}{t'},\pd{\phi^{k'}}{x^{i'}}\big)}{\eps} \bigg|_{\eps=0} 
\\
=& \Bigg(\pd{\L}{t'}\tau + \pd{\L}{x^{i'}} \xi^i + \pd{\L}{\phi^{k'}}\varphi^k + \pd{\L}{\pd{\phi^{k'}}{t'}} \dd{}{\eps} \pd{\phi^{k'}}{t'} 
+ \pd{\L}{\pd{\phi^{k'}}{x^{i'}}} \dd{}{\eps} \pd{\phi^{k'}}{x^{i'}} \Bigg)\bigg|_{\eps=0} 
\\
=&\Bigg(\pd{\L}{(t+\eps \tau)}\tau 
+ \pd{\L}{(x^i+\eps\xi^i)} \xi^i 
+ \pd{\L}{(\phi^k+\eps\varphi^k)}\varphi^k \Bigg)\bigg|_{\eps=0}
+\\
& + \pd{\L}{\Dt{(\phi^k)}}\Big( \Dt{(\varphi^k)} - \Dt{(\phi^k)} \Dt \tau \Big)
+\pd{\L}{\phi^k_{,i}}\Big( \varphi^k_{,j} \delta^j_i - \phi^k_{,j} \xi^j_{,i} \Big)
\\
=& \pd{\L}{t}\tau
+\pd{\L}{x^i}\xi^i
+\pd{\L}{\phi^k}\varphi^k
+ \pd{\L}{\Dt{(\phi^k)}}\Big( \Dt{(\varphi^k)} - \Dt{(\phi^k)} \Dt \tau \Big)
+\pd{\L}{\phi^k_{,i}}\Big( \varphi^k_{,i} - \phi^k_{,j} \xi^j_{,i} \Big) \ .
\alend
Now by inserting the latter into Eq.\,\eqref{inv.Lag}, we find the \textsc{Rund--Trautman} identity:
\begal
&\tau \pd{\L}{t} + \xi^i \pd{\L}{x^i} + \varphi^k \pd{\L}{\phi^k} 
+ \Dt{\tau} \bigg(\L - \Dt{(\phi^k)} \pd{\L}{\Dt{(\phi^k)}} \bigg)
+ \xi^j_{,i} \bigg( \L \delta^i_j - \phi^k_{,j} \pd{\L}{\phi^k_{,i}} \bigg)
\\
&+\Dt{(\varphi^k)} \pd{\L}{\Dt{(\phi^k)}}
+\varphi^k_{,i} \pd{\L}{\phi^k_{,i}} = \mp F_k \varphi^k \ .
\label{rundtrautman}
\alend
This \textsc{Rund--Trautman} identity \cite{rund1966hamilton,trautman1967noether,rund1972direct} is general. In the following, we analyze this identity term-by-term and discuss some simplifications leading to well-known scenarios:
\begin{itemize}
\item First term: When \textsc{Lagrange}an does not depend on time, $\p\L/\p t=0$, and the right-hand side vanishes, $F_k=0$, then arbitrary transformations in time are allowed and $\L$ is conserved. This property is known as ``constant energy.''
\item Second term: In the case of homogeneity---\textsc{Lagrange}an is constant in space, $x^k$, thus, the second term vanishes---arbitrary transformation in space is allowed in the case of vanishing right-hand side, $F_k=0$. The simple example is a free motion of a rigid body.
\item Third term: If \textsc{Lagrange}an does not depend on primitive fields, $\phi^k$, leading to, $\p \L/\p\phi^k = 0$, then no supply or volumetric terms apply. In mechanics, supply is because of gravitational or electromagnetic fields.
\item Fourth and fifth terms: Whenever the generator of time transformation depends on time or the generator of space transformation on space, we have a term called energy-momentum tensor if time and space are written together. We discuss these terms in the following in more detail. In mechanics, this term is often simplified and not discussed.
\item The terms $\p \L/\p\Dt{(\phi^k)}$ and $\p \L/\p \phi^k_{,i}$ are called the conjugated momenta, however, in case of fields this name may be misleading.
\end{itemize}
The \textsc{Rund--Trautman} identity is an extension to the classical \textsc{Noether}ian approach, we may claim that this identity is one step before obtaining ``conservation laws.'' Especially the fourth term is important to notice: multiplied by a minus, it is often introduced as \textsc{Hamilton}ian of the system
\begeq
H = -\L + \Dt{(\phi^k)} \pd{\L}{\Dt{(\phi^k)}} \ .
\label{hamiltonian00}
\eqend
Herein, we see that the term is motivated by the \textsc{Rund--Trautman} identity. The latter definition for the \textsc{Hamilton}ian is numerically equal to the canonical \textsc{Hamilton}ian in a \textsc{Lagrange}an formulation \cite{kobe2013noether}. We emphasize that these concepts of \textsc{Lagrange}an and \textsc{Hamilton}ian are used for systems with $t$ as the only independent variable. We continue the \textsc{Noether}ian formalism in the following with time and space as independent variables. First, we combine all of the independent variables together as a set $a^\nu=\{t,x^1,x^2,x^3\}$. We may even think of $\nu=0,1,2,3$ in order to have $a^0=t$ and $a^1=x^1$, $a^2=x^2$, $a^3=x^3$, for a simpler analogy. Second, the linear transformation is rewritten, $a^{\nu'}=a^\nu+\eps \alpha^\nu$. Third, we write the \textsc{Rund--Trautman} identity once more for $\L\big( a^\nu ; \phi^k(a^\nu) , \phi^k_{,\mu} (a^\nu) \big)$ in this notation,
\begeq
\alpha^\nu \L_{,\nu} + \varphi^k \pd{\L}{\phi^k} + \alpha^\mu_{,\nu} \Big( \L \delta^\nu_\mu - \phi^k_{,\mu} \pd{\L}{\phi^k_{,\nu}} \Big) + \varphi^k_{,\mu} \pd{\L}{\phi^k_{,\mu}}=  \mp F_k \varphi^k \ ,
\label{rundtrautman2}
\eqend
where we use $()_{,\nu}=\p()/\p a^\nu$. Now we define the energy-momentum tensor just by rewriting the third term in the latter,
\begeq
\Ha^{\ \nu}_{\mu}=-\L \delta^\nu_\mu + \phi^k_{,\mu} \pd{\L}{\phi^k_{,\nu}} \ .
\label{hamiltonian}
\eqend
We observe that the \textsc{Lagrange} function, $\L$, and its invariance is more fundamental than a theory based on the \textsc{Hamilton} function, $H=\Ha^{\ 0}_{0}$, because it fails to have an invariance property in general. Indeed the ``energy-momentum'' tensor in Eq.\,\eqref{hamiltonian} is a generalization of the \textsc{Hamilton}ian in Eq.\,\eqref{hamiltonian00}. The energy-momentum tensor is composed of \textsc{Hamilton}ian in its time (scalar) part, $H$, and \textsc{Eshelby} stress tensor in its space part $\Ha^{\ i}_{j}$. The term $\PP_k^{\ \nu} = \p \L/\p \phi^k_{,\mu}$ is called a canonical momentum in the case of the rigid body motion, where $\L$ incorporates the kinetic energy without  deformation energy. We prefer to call it a conjugate term, for example, in elasticity, stress is energetic conjugate of strain that is the space derivative of the primitive variable, which is displacement. In this analogy, the space part of the energy-momentum tensor, $\Ha^{\ i}_{j}$, is often introduced by a so-called \textsc{Legendre} transformation. Herein, we realize that the same term is directly generated by the \textsc{Noether} formalism. The result has been obtained in \cite{eckart1960}; however, the condition of functional $\J$ being extremal has also been used. The procedure herein is different and less restrictive since the energy-momentum tensor asserts only the invariance of the \textsc{Lagrange}an but not its density. The invariance and extremality are separate properties, until now, we have only used invariance. 

\section{Conservation laws} \label{sect:conservation}

We will derive conservation laws in two steps, first, we obtain the so-called \textsc{Euler--Lagrange} equations, second, we use them in the \textsc{Rund--Trautman} identity in order to obtain the conservation laws. They are the balance laws derived by using the formalism herein. We begin with $\alpha^\nu\equiv 0$, where this restriction means that we neglect, for the moment, any shift of the independent variables, i.e. no time and space variation. First and third terms vanish in Eq.\,\eqref{rundtrautman2} and we obtain
\begeq
\varphi^k \pd{\L}{\phi^k} + \varphi^k_{,\mu} \pd{\L}{\phi^k_{,\mu}} = \mp F_k \varphi^k \ .
\eqend
This identity is rewritten in an integral form over a space-time domain $\Om$ and integrated by parts,
\begeq\label{integral.form.to.local.form}
\int_\Om \Big( \varphi^k \pd{\L}{\phi^k} + \varphi^k_{,\mu} \pd{\L}{\phi^k_{,\mu}} \pm F_k \varphi^k \Big) \d \Sigma =0 \ , \\
\int_\Om \Big( \varphi^k \pd{\L}{\phi^k} - \varphi^k \Big(\pd{\L}{\phi^k_{,\mu}}\Big)_{,\mu} \pm F_k \varphi^k \Big) \d \Sigma + \oint_{\p\Om} \varphi^k \pd{\L}{\phi^k_{,\mu}} \d S_\mu =0 \ ,
\eqend
where the surface integral, $\d S_\mu = n_\mu \d S $, is computed on the boundaries with surface normal, $n_\mu$, this normal comprises space---on boundaries of the continuum body, $n_i$ is known as the outer normal---and time (initial and end time). We are interested in a differential equation within the domain, in other words, values at boundaries are given; no variation is needed, $\varphi^k=0 \ , \ \forall a^\mu \in \p\Om$. Thus the last integral vanishes and the well-known \textsc{Euler--Lagrange} equations appear
\begeq
\pd{\L}{\phi^k} - \Big(\pd{\L}{\phi^k_{,\mu}}\Big)_{,\mu} = \mp F_k \ .
\label{euler-lagrange}
\eqend
This equation is well-known, for example given as the integral \textsc{Lagrange--d'Alembert} principle in \cite[Definition 7.8.4]{marsden2013introduction} for  discrete systems. We assert it herein as an additional condition to the \textsc{Rund--Trautman} identity. It is also called an extremal principle, since in Eq.\,\eqref{variation.assertion}, the directional derivative vanishes, i.e. the condition, $\L' J - \L \pm \del A$, is evaluated at its extremum (minimum, maximum, or saddle point). We realize that for a specific case of no space and time translation, all methods are identical, which we will discuss further in an application. Often, the right-hand side is neglected in fields, we refer to \cite{bersani2020lagrangian} for a connection of the right-hand side with \textsc{Rayleigh} dissipation function. We emphasize that we have neglected a reference frame evolution that is an irreversible phenomenon; for an example, we refer to Appendix \ref{appendix} with an application. This \textsc{Euler--Lagrange} equation is obtained by using an invariant but it does not mean that the \textsc{Euler--Lagrange} equation remains the same under coordinate transformations. Their form changes; balance of momentum is a typical example in mechanics, under a proper coordinate transformation, additional terms emerge. If there is a coordinate transformation leaving the \textsc{Euler--Lagrange} equation unchanged then this transformation is called \textit{invariant transformation}. Herein, we use a formalism for arbitrary transformations. 

As the right-hand side is a (non-conservative) force, it is rank 1. Analogous to the aforementioned case, without introducing any assumption or reduction, we may deduce a rank 1 tensor from a rank 2 tensor by (space and time) derivative
\begeq
\pd{\L}{\phi^k} - \PP_{k,\mu}^{\ \mu} = S_{k,\mu}^{\ \mu} \comma \PP_{k}^{\ \mu} = \pd{\L}{\phi^k_{,\mu}} \ .
\label{euler-lagrange2}
\eqend
For the derivation, we have used an oblique but not curvilinear coordinate system (we refer to \cite[\textsection~43]{pauli2}, \cite{dirac} for a generic derivation with \textsc{Christoffel} symbols). Now we start with Eq.\,\eqref{hamiltonian} and obtain, by using $\L=\L(a^\mu; \phi^k(a^\mu), \phi^k_{,\nu}(a^\mu))$ and chain rule,
\begeq
\Ha_{\nu,\rho}^{\ \rho} = \Big( - \L \delta_\nu^\rho + \phi^k_{,\nu} \PP_k^{\ \rho}  \Big) _{,\rho}
=-\L_{,\nu} -\phi^k_{,\nu} \pd{\L}{\phi^k} - \phi^k_{,\mu\nu} \PP_k^{\ \mu} + \Big( \phi^k_{,\nu} \PP_k^{\ \rho}  \Big) _{,\rho}
=-\L_{,\nu} -\phi^k_{,\nu} \pd{\L}{\phi^k} + \phi^k_{,\nu} \PP_{k,\rho}^{\ \rho} \ .
\eqend
By inserting the latter in the \textsc{Rund--Trautman} identity in Eq.\,\eqref{rundtrautman2}, we acquire
\begeq
\alpha^\nu \L_{,\nu} + \varphi^k \pd{\L}{\phi^k} - \alpha^\mu_{,\nu} \Ha^{\ \nu}_{\mu} +  \varphi^k_{,\mu} \PP_k^{\ \mu}  = \varphi^k S_{k ,\mu}^{\ \mu}  \ ,
\\
\alpha^\nu \bigg( \L_{,\nu} + \Ha_{\nu,\rho}^{\ \rho} \bigg)
+ \varphi^k \pd{\L}{\phi^k}
-\bigg(\alpha^\mu \Ha_\mu^{\ \nu} \bigg)_{,\nu} 
+\bigg( \varphi^k \PP_k^{\ \mu} \bigg)_{,\mu} - \varphi^k \PP_{k,\mu}^{\ \mu} 
= \varphi^k S_{k ,\mu}^{\ \mu} \ , \\
-\alpha^\nu \phi^k_{,\nu} \bigg( \pd{\L}{\phi^k} - \PP_{k,\rho}^{\ \rho} \bigg)
+ \varphi^k \bigg( \pd{\L}{\phi^k} - \PP_{k,\mu}^{\ \mu} - S_{k ,\mu}^{\ \mu} \bigg)
+\bigg( \varphi^k \PP_k^{\ \nu} - \alpha^\mu \Ha_\mu^{\ \nu} \bigg)_{,\nu}  = 0
\ .
\eqend
By assuming that \textsc{Euler--Lagrange} equations hold, we insert Eq.\,\eqref{euler-lagrange2} and obtain
\begeq
\bigg( \varphi^k \PP_k^{\ \nu} - \alpha^\mu \Ha_\mu^{\ \nu} \bigg)_{,\nu} = \alpha^\nu \phi^k_{,\nu} S_{k ,\mu}^{\ \mu} \ .
\eqend
By renaming the left-hand side as \textsc{Noether}'s current, $\Curr^\mu$, we obtain the corresponding equation:
\begeq
\Curr^\mu_{,\mu}=\alpha^\nu \phi^k_{,\nu} S_{k ,\mu}^{\ \mu} \comma 
\Curr^\mu=\varphi^k \pd{\L}{\phi^k_{,\mu}} - \alpha^\nu \Ha^{\ \mu}_{\nu} \ .
\label{conservation.law}
\eqend
In the case of vanishing right-hand side, $S_{k ,\mu}^{\ \mu}=\pm F_k=0$, the latter ``balance'' equations are called conservation laws, $\Curr^\mu_{,\mu}=0$. 

\section{Elastodynamics}\label{sect:first.order}

In order to demonstrate the meaning of conservation laws, we give an example in elastodynamics. The primitive variables, $\phi^k$, are the components of the displacement field, $u_1$, $u_2$, $u_3$, expressed in Cartesian coordinates. Reference frame is not evolving such that we have a reversible process. Let us use the following \textsc{Lagrange}an density:
\begeq
\L = \frac12 \tilde C_{i\mu k \nu} u_{i,\mu} u_{k,\nu} \comma
\tilde C_{i\mu k \nu} = \begin{cases}
\rho_\RRef \delta_{ik} & \text{  if  } \mu=0 , \nu=0 \ , \\
0 &  \text{  if  } \mu=0 , \nu\neq 0 \text{  or  } \mu\neq 0 , \nu=0 \ , \\
-C_{ijkl} & \text{  if  } \mu=j , \nu=l \ ,
\end{cases}
\label{lagrange.elastodynamics}
\eqend
where the stiffness tensor, $C_{ijkl}$, is given for the corresponding material. Mass density, $\rho_\RRef$, is defined on the reference frame, thus, it is constant in time. In the case of linear and isotropic material, the stiffness tensor reads $C_{ijkl} = \lambda \delta_{ij} \delta_{kl} + \mu \delta_{ik} \delta_{jl} + \mu \delta_{il} \delta_{jk}$, where the so-called \textsc{Lame} parameters are given by engineering constants; \textsc{Young}'s modulus, $E$, and \textsc{Poisson}'s ratio, $\nu$, as follows:
\begeq
\lambda=\frac{E \nu}{(1+\nu)(1-2\nu)} \comma
\mu = \frac{E}{2(1+\nu)} \ .
\eqend
We separate time, $\mu=0$, and space, $\mu=\{1,2,3\}=i$, for a direct analogy with linear elasticity theory. Hence, the aforementioned \textsc{Lagrange}an density in space and time reads 
\begeq
\L = \frac12 \rho_\RRef \Dt u_i \Dt u_i - w  \comma 
w = \frac12 u_{i,j} C_{ijkl} u_{k,l} \comma
\eqend
with the deformation (or stored) energy density, $w$ in J/m$^3$. Instead of starting with Eq.\eqref{lagrange.elastodynamics}, we may introduce \textsc{Lagrange}an density as kinetic energy density minus deformation energy density. Albeit not immediately obvious, we use a linear strain measure, $E_{ij} = 1/2(u_{i,j}+u_{j,i})$, thus, the stored energy density, $w=1/2 E_{ij} C_{ijkl} E_{kl}$, is objective and reduces to the quadratic energy, $w = 1/2 u_{i,j} C_{ijkl} u_{k,l}$, effected by the minor symmetries, $C_{ijkl}=C_{jikl}=C_{ijlk}$. The formulation is analogous for finite strain formulation, for demonstrating this, we use a stress measure in the following. By choosing the deformation energy differently, for example as known in hyperelasticity, stress is calculated by its derivative with respect to displacement gradient. 

In the case of vanishing viscous effects, $F_k=0$, we rewrite the conservation laws in Eq.\,\eqref{conservation.law}, as follows: 
\begeq
\Dt \Curr_0 + \Curr_{i,i} = 0 \comma
\Curr_0 = \varphi_k \pd{\L}{\Dt u_k} - \tau \Ha_{00} - \xi_i \Ha_{i0} \comma
\Curr_i = \varphi_k \pd{\L}{u_{k,i} } - \tau \Ha_{0i} - \xi_j \Ha_{ji} \ .
\eqend
The whole formulation is in material frame, although for simplicity, we ignore the difference between reference and current frame. The energy-momentum tensor reads with a time part called \textsc{Hamilton}ian and a space part called \textsc{Eshelby} stress tensor
\begeq
\Ha_{\alpha \beta}
=-\L \delta_{\alpha \beta} + u_{k,\alpha} \pd{\L}{u_{k,\beta}}  \ .
\eqend
We use its counterpart in space and time
\begal
\Ha_{0 0} =& - \L + \Dt u_k \pd{\L}{\Dt u_k}  = - \frac12 \rho_\RRef \Dt u_i \Dt u_i + w + \Dt u_k \rho_\RRef \Dt u_k 
= \frac12 \rho_\RRef \Dt u_i \Dt u_i + w   \ , \\
\Ha_{0 i} =& \Dt u_k \pd{\L}{u_{k,i}}  = \Dt u_k \PP_{ki} \comma \PP_{ki} = \pd{\L}{u_{k,i}} = -\pd{w}{u_{k,i}} = - C_{kilm} u_{l,m} = - \sigma_{ki} \ , \\
\Ha_{i 0} =&  u_{k,i} \pd{\L}{\Dt u_k}  = u_{k,i} \rho_\RRef \Dt u_k  \ , \\
\Ha_{i j} =& -\L \delta_{ij} + u_{k,i} \pd{\L}{u_{k,j}}  = -\frac12 \rho_\RRef \Dt u_k \Dt u_k \delta_{ij} + w \delta_{ij} + u_{k,i} \PP_{kj}  
\ .
\alend
\textsc{Noether} currents become
\begal
\Curr_0 =& \varphi_k \rho_\RRef \Dt u_k - \frac{\tau}{2} \rho_\RRef \Dt u_i \Dt u_i - \tau w   - \xi_i u_{k,i} \rho_\RRef \Dt u_k \comma \\
\Curr_i =& \varphi_k \PP_{ki} - \tau \Dt u_k \PP_{ki}  + \xi_j \Big( \frac12 \rho_\RRef \Dt u_k \Dt u_k \delta_{ji} - w \delta_{ji}  - u_{k,j} \PP_{ki}  \Big) \ .
\alend
Technically, $\PP_{ij}$ is the minus nominal stress or minus transpose of \textsc{Piola} stress, indeed, in small strain assumption that corresponds to the minus \textsc{Cauchy} stress. The conservation laws,  
\begal
0 =& \Dt \Curr_0 + \Curr_{i,i} 
\\
0=& \Dt{\Big( \varphi_k \rho_\RRef \Dt u_k - \frac{\tau}{2} \rho_\RRef \Dt u_i \Dt u_i - \tau w - \xi_i u_{k,i} \rho_\RRef \Dt u_k \Big)}
\\
&+ \bigg( \varphi_k \PP_{ki} - \tau \Dt u_k \PP_{ki}  + \xi_j \Big( \frac12 \rho_\RRef \Dt u_k \Dt u_k \delta_{ji} - w \delta_{ji} - u_{k,j} \PP_{ki} \Big) \bigg)_{,i} 
\alend
may be rewritten as follows, as they hold for arbitrary transformations,
\begal
0=& 
\varphi_k \bigg( \rho_\RRef \DDt u_k + \PP_{ki,i} \bigg)
-\tau \bigg( \Dt{\Big( \frac12 \rho_\RRef \Dt u_i \Dt u_i +  w \Big)}  + \big( \Dt u_k \PP{ki}\big)_{,i} \bigg)
\\
&+\xi_j  \bigg( -  \Dt{\big(\rho_\RRef \Dt u_k u_{k,j} \big)} + \Big( \frac12 \rho_\RRef \Dt u_k \Dt u_k \delta_{ji} - w \delta_{ji} - u_{k,j} \PP_{ki}\Big)_{,i} \bigg)
\\
&+\Dt \varphi_k  \rho_\RRef \Dt u_k 
-\Dt \tau  \bigg(\frac12 \rho_\RRef \Dt u_i \Dt u_i + w \bigg)
-\Dt \xi_j   \rho_\RRef \Dt u_k u_{k,j} 
+\varphi_{k,i} \PP_{ki} 
\\
&-\tau_{,i}  \Dt u_k \PP_{ki} 
+\xi_{j,i}  \bigg( \frac12 \rho_\RRef \Dt u_k \Dt u_k \delta_{ji} - w \delta_{ji} - u_{k,j} \PP_{ki} \bigg) \ .
\label{balance.laws}
\alend
We stress that the transformations in Eq.\,\eqref{trafo.arbitrary} may be chosen in such a way that the above equation holds. In this way, it is possible to formulate an inverse problem and search for possible transformations by solving so-called \textsc{Killing} equations from the latter. A direct problem is to test different transformations and find out the consequence of invariance and extremality of $\J$. The former has brought us the \textsc{Rund--Trautman} identity and the latter \textsc{Euler--Lagrange} equations. By using both of them, we have obtained \textsc{Noether}'s currents leading to Eq.\,\eqref{balance.laws}. If we examine a transformation in displacement field, $\varphi_k=\text{const}$, and insert it into the latter, we obtain the balance of linear momentum, 
\begeq
0=\varphi_k \bigg( \rho_\RRef \DDt u_k + \PP_{ki,i} \bigg) \ .
\eqend
We emphasize that $\PP_{ij}=-\sigma_{ji}$ in small deformation assumption. Since the transformation is constant in space and time, such a displacement is called a rigid body motion. We may write the result on material frame for a domain $\B$, with its closure (smooth boundary) $\p\B$, after applying \textsc{Gauss--Ostrogradskiy}'s theorem,
\begeq
\Dt{\bigg( \int_\B \rho_\RRef \Dt u_k \d V \bigg)} = \int_{\p\B} \bar t_k \d A \comma \bar t_k = - \PP_{ki} n_i = \sigma_{ik} n_i  \ ,
\eqend
where the traction vector $\bar t_i$ is defined on the boundary. Without using \textsc{Cauchy}'s tetrahedron argumentation or usual balance equations' argumentation, we obtain the result as a consequence of the transformation rule. The justification is obvious that we search for laws holding under rigid body translations. It is possible to call that a chosen symmetry, $\varphi_k=\text{const}$, generates a balance law. This result is not an additional balance law, since we have to satisfy \textsc{Euler--Lagrange} equations,
\begeq
\pd{\L}{u_k} - \PP_{ki,i} - \Dt \PP_{k0}= 0 \comma 
\pd{\L}{u_k} = 0 \comma 
\PP_{k0} = \rho_\RRef \Dt u_k \comma 
\PP_{ki} = -\pd{w}{u_{k,i}} = -\sigma_{ki} \ , 
\eqend
leading to the same governing equation. This consequence is in fact the aforementioned relation that \textsc{Rund--Trautman} identity reduces to the \textsc{Euler--Lagrange} equations for the case of no space and time variations. More interestingly, now we may easily examine other transformation rules in order to acquire additional governing equations. 

Analogously, we may examine a time translation, $\tau=\text{const}$, in order to obtain the balance of energy with the total specific (per mass) energy, $e$ in J/kg,
\begeq
\Dt{\bigg( \rho_\RRef e \d V \bigg)} = \int_{\p\B} \Dt u_i \bar t_i \d A \comma e = \frac12 \Dt u_i \Dt u_i +  \frac{w}{\rho_\RRef}  \ .
\eqend 
A \textsc{Galilei}an transformation, for example $\xi_i=\text{const}$, reads
\begeq
\Dt{\bigg( \int_\B \rho_\RRef \Dt u_k u_{k,j} \d V \bigg)} = \int_{\p\B} \big( n_j \L + u_{k,j} \bar t_k \big) \d A   \ .
\eqend 
This new balance law reduces to the well-known $J$-integral with the \textsc{Eshelby} stress tensor,
\begeq
0 = \int_{\p\B} \big( - n_j w + u_{k,j} \bar t_k \big) \d A   \ ,
\eqend 
for the stationary case. If the integral is taken around a crack tip, the value of this integral is seen as an energy release rate for forming a discontinuity (propagating a crack) \cite{eshelby1956continuum, knowles1973asymptotic, maugin2020material}. Herein, its dynamical counterpart is acquired by the \textsc{Noether} formalism, so its interpretation and use are more obvious. In general, we may skip this balance law and hope that it is fulfilled, but actually, adding an additional restriction is a remedy in numerical accuracy related problems \cite{gurses2009computational}. More different transformations may be examined, for example scaling or rotation (leading to the balance of angular momentum) is studied in \cite{fletcher1976conservation}. A generalization of this formalism for thermoelasticity is possible as in \cite{kalpakides2004canonical}. 

We have obtained two governing equations to be fulfilled in an isothermal case. One is the term multiplied by $\varphi_k$ and the other is the term multiplied by $\xi_i$.  

\section{Generalized continua}\label{sect:second.order}

Without repeating all the analysis, we now address the case if the \textsc{Lagrange}an depends also on the second derivative $\L\big(a^\nu;\phi^k(a^\nu),\phi^k_{,\mu}(a^\nu),\phi^k_{,\mu\gamma}(a^\nu)\big)$. As the transformation is still linear, we have extra terms in Eq.\,\eqref{variational.part.formulation.term.one} appearing for the term multiplied by $\alpha^\mu$ leading to the  \textsc{Rund--Trautman} identity for generalized continua,
\begeq
\alpha^\nu \L_{,\nu} + \varphi^k \pd{\L}{\phi^k} 
+ \alpha^\mu_{,\nu} \Big( \L \delta^\nu_\mu - \phi^k_{,\mu} \pd{\L}{\phi^k_{,\nu}} - \phi^k_{,\gamma\mu} \pd{\L}{\phi^k_{,\gamma\nu}} \Big) 
+ \varphi^k_{,\mu} \pd{\L}{\phi^k_{,\mu}} + \varphi^k_{,\gamma\mu} \pd{\L}{\phi^k_{,\gamma\mu}} =   \varphi^k S_{k ,\mu}^{\ \mu}\ ,
\label{rundtrautman.generalized}
\eqend
Therefore, the energy-momentum tensor is renewed
\begeq
\Ha^{\ \nu}_{\mu}=-\L \delta^\nu_\mu + \phi^k_{,\mu} \pd{\L}{\phi^k_{,\nu}} + \phi^k_{,\gamma\mu} \pd{\L}{\phi^k_{,\gamma\nu}} \ .
\label{hamiltonian.generalized}
\eqend
With the same integral form as in Eq.\,\eqref{integral.form.to.local.form} and in this round twice-integrating by parts, we obtain \textsc{Euler--Lagrange} equations in generalized continua
\begeq
\pd{\L}{\phi^k} - \Big(\pd{\L}{\phi^k_{,\mu}}\Big)_{,\mu} + \Big(\pd{\L}{\phi^k_{,\gamma\mu}}\Big)_{,\gamma\mu}  = S_{k ,\mu}^{\ \mu} \ .
\label{euler-lagrange.generalized}
\eqend
For a direct analogy, we utilize the same notation as in the previous section with an extension,
\begeq
\PP_{k}^{\ \mu} = \pd{\L}{\phi^k_{,\mu}} \comma
\RR_k^{\gamma\mu} = \pd{\L}{\phi^k_{,\gamma\mu}} \ ,
\eqend
and rewrite the \textsc{Euler--Lagrange} equations:
\begeq
\pd{\L}{\phi^k} - \PP_{k \ ,\mu}^{\ \mu} + \RR_{k \ ,\gamma\mu}^{\gamma\mu}   = S_{k ,\mu}^{\ \mu} \ .
\label{euler-lagrange.generalized2}
\eqend
We repeat the same procedure
\begal
\Ha_{\nu,\rho}^{\ \rho} =& \Big( - \L \delta_\nu^\rho + \phi^k_{,\nu} \PP_k^{\ \rho} + \phi^k_{,\gamma\nu} \RR_k^{\gamma\rho} \Big) _{,\rho}
\\=&-\L_{,\nu} -\phi^k_{,\nu} \pd{\L}{\phi^k} - \phi^k_{,\mu\nu} \PP_k^{\ \mu} - \phi^k_{,\gamma\mu\nu} \RR_k^{\gamma\mu} + \Big( \phi^k_{,\nu} \PP_k^{\ \rho} + \phi^k_{,\gamma\nu} \RR_k^{\ \gamma\rho}  \Big) _{,\rho}
\\=&-\L_{,\nu} -\phi^k_{,\nu} \pd{\L}{\phi^k} + \phi^k_{,\nu} \PP_{k,\rho}^{\ \rho} + \phi^k_{,\gamma\nu} \RR_{k \ \ ,\rho}^{\gamma\rho}
\\=&-\L_{,\nu} -\phi^k_{,\nu} \pd{\L}{\phi^k} + \phi^k_{,\nu} \PP_{k,\rho}^{\ \rho} + \Big( \phi^k_{,\nu} \RR_{k \ \ ,\rho}^{\gamma\rho} \Big)_{,\gamma} - \phi^k_{,\nu} \RR_{k \ \ ,\gamma\rho}^{\gamma\rho}\ .
\alend
By using the latter in the \textsc{Rund--Trautman} identity in Eq.\,\eqref{rundtrautman.generalized}, we obtain
\begal
& \alpha^\nu \L_{,\nu} + \varphi^k \pd{\L}{\phi^k} - \alpha^\mu_{,\nu} \Ha^{\ \nu}_{\mu} +  \varphi^k_{,\mu} \PP_k^{\ \mu}  + \varphi^k_{,\gamma\mu} \RR_k^{\gamma\mu} = \varphi^k S_{k ,\mu}^{\ \mu}  
\ , \\
& \alpha^\nu \bigg( \L_{,\nu} + \Ha_{\nu,\rho}^{\ \rho} \bigg)
+ \varphi^k \pd{\L}{\phi^k}
-\bigg(\alpha^\mu \Ha_\mu^{\ \nu} \bigg)_{,\nu} 
+\bigg( \varphi^k \PP_k^{\ \mu} \bigg)_{,\mu} - \varphi^k \PP_{k,\mu}^{\ \mu} 
+\\
&\quad +\bigg( \varphi^k_{,\gamma} \RR_k^{\gamma\mu} \bigg)_{,\mu} 
- \varphi^k_{,\gamma} \RR_{k\ ,\mu}^{\gamma\mu} 
= \varphi^k S_{k ,\mu}^{\ \mu} 
\ , \\
& \alpha^\nu \bigg( \L_{,\nu} + \Ha_{\nu,\rho}^{\ \rho} \bigg)
+ \varphi^k \pd{\L}{\phi^k}
-\bigg(\alpha^\mu \Ha_\mu^{\ \nu} \bigg)_{,\nu} 
+\bigg( \varphi^k \PP_k^{\ \mu} \bigg)_{,\mu} - \varphi^k \PP_{k,\mu}^{\ \mu} 
+ \\
& \quad +\bigg( \varphi^k_{,\gamma} \RR_k^{\gamma\mu} \bigg)_{,\mu} 
-\Big( \varphi^k \RR_{k\ ,\mu}^{\gamma\mu} \Big)_{,\gamma} + \varphi^k \RR_{k\ ,\gamma\mu}^{\gamma\mu}
= \varphi^k S_{k ,\mu}^{\ \mu} 
\alend
which is rewritten, as follows:
\begal
& -\alpha^\nu \phi^k_{,\nu} \bigg( \pd{\L}{\phi^k} - \PP_{k,\rho}^{\ \rho} + \RR_{k \ \ ,\gamma\rho}^{\gamma\rho} \bigg)
+ \varphi^k \bigg( \pd{\L}{\phi^k} - \PP_{k,\mu}^{\ \mu} + \RR_{k\ ,\gamma\mu}^{\gamma\mu} - S_{k ,\mu}^{\ \mu} \bigg)
+\\
& \quad +\bigg( \varphi^k \big( \PP_k^{\ \nu} - \RR_{k\ ,\mu}^{\nu\mu} \big) + \varphi^k_{,\gamma} \RR_k^{\gamma\nu} - \alpha^\mu \Ha_\mu^{\ \nu} \bigg)_{,\nu}  = 0
\ .
\alend
Into the latter, we insert the generalized \textsc{Euler--Lagrange} equations in Eq.\,\eqref{euler-lagrange.generalized2} and obtain generalized \textsc{Noether}'s current,
\begeq
\bigg( \varphi^k \big( \PP_k^{\ \nu} - \RR_{k\ ,\mu}^{\nu\mu} \big) + \varphi^k_{,\gamma} \RR_k^{\gamma\nu} - \alpha^\mu \Ha_\mu^{\ \nu} \bigg)_{,\nu} = \alpha^\nu \phi^k_{,\nu} S_{k ,\mu}^{\ \mu} 
\ , \\
\Curr^\mu_{,\mu} = \alpha^\nu \phi^k_{,\nu} S_{k ,\mu}^{\ \mu} \comma 
\Curr^\mu = \varphi^k \bigg( \pd{\L}{\phi^k_{,\mu}} - \Big( \pd{\L}{\phi^k_{,\gamma\mu}} \Big)_{,\gamma} \bigg) 
+ \varphi^k_{,\gamma} \pd{\L}{\phi^k_{,\gamma\mu}}
- \alpha^\nu \Ha^{\ \mu}_{\nu} \ .
\label{conservation.law.generalized}
\eqend
We follow the same guidelines and generalize to higher order continua for metamaterials. The primitive variable is again ``only'' the displacement, $\phi^k=u_k$, expressed in Cartesian coordinates. But now the second derivative plays a role as well, so we use the following \textsc{Lagrange}an density:
\begal
\L =& \frac12 \tilde C_{i\mu k \nu} u_{i,\mu} u_{k,\nu} + \frac12 \tilde D_{i\mu\gamma l \nu\eta} u_{i,\mu\gamma} u_{l,\nu\eta} + \tilde G_{i\mu k \nu\eta} u_{i,\mu} u_{k,\nu\eta} 
\ , \\
\tilde C_{i\mu k \nu} =& 
\begin{cases}
\rho_\RRef \delta_{ik} & \text{  if  } \mu=\nu=0 \ , \\
-C_{ijkl} & \text{  if  } \mu=j , \nu=l \ , \\
0 &  \text{  else  }  \ , 
\end{cases}
\ , \\
\tilde D_{i\mu\gamma l \nu\eta} =& 
\begin{cases}
\rho_\RRef \tau_\RRef^2 \delta_{il} & \text{  if  } \mu=\nu=0 , \gamma=\eta=0  \ , \\
\rho_\RRef d_\RRef^2 \delta_{il} \delta_{jk} & \text{  if  } \mu=\nu=0 , \gamma=j , \eta= k \ , \\
-D_{ijklmn} & \text{  if  } \mu=j , \gamma=k , \nu=m , \eta=n \ , \\
0 &  \text{  else  }  \ , 
\end{cases}
\ , \\
\tilde G_{i\mu k \nu\eta} =& 
\begin{cases}
-G_{ijklm} & \text{  if  } \mu=j , \nu=l , \eta=m \ , \\
0 &  \text{  else  }  \ , 
\end{cases}
\alend
where rank 4, 5, 6 tensors, $C_{ijkl}$, $D_{ijklmn}$, $G_{ijklm}$, are given for the corresponding metamaterial. Their measurement seems to be challenging \cite{049,muller2020experimental,askes2011gradient}, yet there exist different homogenization methods \cite{auffray2010strain,tran2012micromechanics,barboura2018establishment, 057,069,solyaev2022self} that calculate these parameters \cite{084,085,yvonnet2020computational,lahbazi2022size,areias2022finite}. Inertia related terms, $\rho_\RRef$, $d_\RRef$, $\tau_\RRef$, are all defined on the reference frame. We redo the same analysis as before and separate time and space, in order to obtain
\begal
\L =& \frac12 \rho_\RRef \big( \Dt u_i \Dt u_i + \tau_\RRef^2 \DDt u_i \DDt u_i + d_\RRef^2 \Dt u_{i,j} \Dt u_{i,j} \big) - w  
\ , \\
w =& \frac12 u_{i,j} C_{ijkl} u_{k,l} + \frac12 u_{i,jk} D_{ijklmn} u_{l,mn} + G_{ijklm} u_{i,j} u_{k,lm} \ .
\alend
In the case of elasticity, $S_{k ,\mu}^{\ \mu}=0$, we acquire the conservation laws in Eq.\,\eqref{conservation.law}, as follows: 
\begal
\Dt \Curr_0 + \Curr_{i,i} =& 0 \ , \\
\Curr_0 =& \varphi_k \bigg( \pd{\L}{\Dt u_k} - \Dt{\Big( \pd{\L}{\DDt u_{k}} \Big)} - \Big( \pd{\L}{\Dt u_{k,i}} \Big)_{,i} \bigg)
+ \Dt\varphi_k \pd{\L}{\DDt u_k} 
+ \varphi_{k,i} \pd{\L}{\Dt u_{k,i}} 
- \tau \Ha_{00} - \xi_i \Ha_{i0} \ , \\
\Curr_i =& \varphi_k \bigg( \pd{\L}{u_{k,i}} - \Dt{\Big( \pd{\L}{\Dt u_{k,i}} \Big)} - \Big( \pd{\L}{u_{k,ij}} \Big)_{,j} \bigg)
+ \Dt\varphi_k \pd{\L}{\Dt u_{k,i}} 
+ \varphi_{k,j} \pd{\L}{u_{k,ij}}
 - \tau \Ha_{0i} - \xi_j \Ha_{ji} \ .
\alend
Since we are interested in the terms multiplied by $\varphi_k$ and $\xi_i$, only the following terms are sought after
\begeq
\varphi_k \Bigg(
\Dt{ \bigg( \pd{\L}{\Dt u_k} - \Dt{\Big( \pd{\L}{\DDt u_{k}} \Big)} - \Big( \pd{\L}{\Dt u_{k,i}} \Big)_{,i} \bigg) }
+ \bigg( \pd{\L}{u_{k,i}} - \Dt{\Big( \pd{\L}{\Dt u_{k,i}} \Big)} - \Big( \pd{\L}{u_{k,ij}} \Big)_{,j} \bigg)_{,i} \Bigg) = 0
\eqend
and
\begeq
-\xi_i \Dt \Ha_{i0} - \xi_j \Ha_{ji,i} = 0 \ , \\
\xi_i \big( \Dt \Ha_{i0} + \Ha_{ij,j} \big) = 0 \ .
\eqend
For the generalized elasticity, from the terms multiplied by $\varphi_k$, we acquire 
\begal
& \Dt{ \bigg( \rho_\RRef \Dt u_k - \Dt{\Big( \rho_\RRef \tau_\RRef^2 \DDt u_k  \Big)} - \Big( \rho_\RRef d_\RRef^2 \Dt u_{k,i} \Big)_{,i} \bigg) }
+ \bigg( -\pd{w}{u_{k,i}} - \Dt{\Big( \rho_\RRef d_\RRef^2 \Dt u_{k,i}  \Big)} + \Big( \pd{w}{u_{k,ij}} \Big)_{,j} \bigg)_{,i}  = 0 \ , \\
& \rho_\RRef \DDt u_k -  \rho_\RRef \tau_\RRef^2 \DDDDt u_k  - \Big( \rho_\RRef d_\RRef^2 \DDt u_{k,i} \Big)_{,i}
+ \bigg( - \big( C_{kijl} u_{j,l} + G_{kijlm} u_{j,lm} \big) 
- \rho_\RRef d_\RRef^2  \DDt u_{k,i} 
+ \\
& \quad + \Big( D_{kijlmn} u_{l,mn} + u_{m,n} G_{mnkij} \Big)_{,j} \bigg)_{,i}  = 0 \ , \\
\alend
In the case of a homogeneous material, where $\rho_\RRef$ is constant in space, and a centro-symmetric metamaterial, $G_{ijklm}=0$, analogous to the previous case, we may use stress $\sigma_{kj} = C_{kjlm}  u_{l,m}$ and double stress $m_{kij} = D_{kijlmn} u_{l,mn}$, in order to obtain
\begeq
\rho_\RRef \DDt u_k 
-  \rho_\RRef \tau_\RRef^2 \DDDDt u_k  
- 2 \rho_\RRef d_\RRef^2 \DDt u_{k,ii} 
- \sigma_{ki,i}  
+  m_{kij,ji}  = 0 \ .
\eqend
For the generalized mechanics, the energy-momentum tensor becomes
\begeq
\Ha_{\alpha \beta}
=-\L \delta_{\alpha \beta} + u_{k,\alpha} \pd{\L}{u_{k,\beta}}  
+ u_{k,\gamma\alpha} \pd{\L}{u_{k,\gamma\beta}}\ ,
\eqend
we write it in terms of space and time,
\begal
\Ha_{0 0} =& - \L + \Dt u_k \pd{\L}{\Dt u_k} + \DDt u_k \pd{\L}{\DDt u_k} + \Dt u_{k,i} \pd{\L}{\Dt u_{k,i}} 
= \frac12 \rho_\RRef \Big( \Dt u_i \Dt u_i + \tau_\RRef^2 \DDt u_i \DDt u_i + d_\RRef^2 \Dt u_{i,j} \Dt u_{i,j} \Big) + w  \ , \\
\Ha_{0 i} =& \Dt u_k \pd{\L}{u_{k,i}} + \DDt u_k \pd{\L}{\Dt u_{k,i}}  + \Dt u_{k,j} \pd{\L}{u_{k,ji}}  
\\ =& - \Dt u_k \Big( C_{kilm} u_{l,m} + G_{kijlm} u_{j,lm} \Big) + \DDt u_k \rho_\RRef d_\RRef^2 \Dt u_{k,i} - \Dt u_{k,j} \Big( D_{kjilmn} u_{l,mn} + u_{l,m} G_{lmkji} \Big) \ , \\
\Ha_{i 0} =&  u_{k,i} \pd{\L}{\Dt u_k} + \Dt u_{k,i} \pd{\L}{\DDt u_k} + u_{k,ji} \pd{\L}{\Dt u_{k,j}}  
= u_{k,i} \rho_\RRef \Dt u_k + \Dt u_{k,i} \rho_\RRef \tau^2_\RRef \DDt u_k + u_{k,ji} \rho_\RRef d_\RRef^2 \Dt u_{k,j}  \ , \\
\Ha_{i j} =& -\L \delta_{ij} + u_{k,i} \pd{\L}{u_{k,j}} + \Dt u_{k,i} \pd{\L}{\Dt u_{k,j}} + u_{k,li} \pd{\L}{u_{k,lj}} 
\\=& -\frac12 \rho_\RRef \Big( \Dt u_k \Dt u_k + \tau_\RRef^2 \DDt u_k \DDt u_k + d_\RRef^2 \Dt u_{k,l} \Dt u_{k,l} \Big) \delta_{ij} + w \delta_{ij} 
- u_{k,i} \Big( C_{kjlm}  u_{l,m} + G_{kjlmn} u_{l,mn} \Big) +
\\ & + \Dt u_{k,i} \rho_\RRef d^2_\RRef \Dt u_{k,j} - u_{k,li} \Big( D_{kljmno} u_{m,no} + u_{m,n} G_{mnklj} \Big)
\ .
\alend
By using the latter, we obtain the generalized $J$-integral
\begeq
\int_\B \Big( \Dt \Ha_{i0} + \Ha_{ij,j} \Big) \d V = 0 \\
\Dt{\bigg( \int_\B \Ha_{i0} \d V \bigg)}= - \int_{\p\B} n_j \Ha_{ij}  \d A \ ,
\eqend
as follows:
\begal 
& \Dt{ \bigg( 
\int_\B \Big( u_{k,i} \rho_\RRef \Dt u_k + \Dt u_{k,i} \rho_\RRef \tau^2_\RRef \DDt u_k + u_{k,ji} \rho_\RRef d_\RRef^2 \Dt u_{k,j} \Big)  \d V
\bigg) }
=
\int_{\p\B}  \bigg( n_i \L
+ \\ 
& \quad + n_j \Big( 
u_{k,i} \Big( C_{kjlm}  u_{l,m} + G_{kjlmn} u_{l,mn} \Big) 
- \Dt u_{k,i} \rho_\RRef d^2_\RRef \Dt u_{k,j} 
+ \\
& \quad + u_{k,li} \Big( D_{kljmno} u_{m,no} + u_{m,n} G_{mnklj} \Big)  \bigg) \d A \ .
\alend
We emphasize that inertial terms arise on the surface integral. Such a result is challenging to obtain without a formal structure as presented herein. In the case of the stationary case, for a centro-symmetric material, $G_{ijklm}=0$, we obtain
\begeq
0=
\int_{\p\B}  \bigg( -n_i w    
+ n_j \Big( 
u_{k,i} C_{kjlm}  u_{l,m}  
+ u_{k,li} D_{kljmno} u_{m,no} \Big) \bigg) \d A \ .
\eqend 
By utilizing $\sigma_{kj} = C_{kjlm}  u_{l,m}$ and $m_{klj} = D_{kljmno} u_{m,no}$ for obtaining traction $\bar t_k = \sigma_{kj} n_j$ and double traction $\bar s_{kl} = m_{klj} n_j$, the generalized $J$-integral reads
\begeq
0=
\int_{\p\B}  \bigg( -n_i w    
+ u_{k,i} \bar t_k  
+ u_{k,li} \bar s_{kl}  \bigg) \d A \ .
\eqend 
Hence, we understand that even a stable crack propagation is steered by not only traction but also double traction. The latter term may be proposed with the help of its structure, but the term from the kinetic energy $\Dt u_{k,i} \rho_\RRef d^2_\RRef \Dt u_{k,j} n_j$ would be missed easily. It is rather difficult to predict its significance in the formulation. We stress that $d_\RRef$ is challenging to measure. In the literature, there are different simplifications for reducing the number of inertial terms \cite{mindlin1968first,altan1997some}. We refer to \cite{082}, for a numerical analysis with an experimental comparison for the role of $d_\RRef$, 
 
\section{Conclusion}

We have revisited the extended \textsc{Noether}'s formalism in continuum mechanics by using tensor algebra and applied directly to elastodynamics. Apart the well-known balance equations, we have observed how the $J$-integral is obtained with the \textsc{Eshelby} stress tensor, which is of importance in modeling damage mechanics. In this way, we understand that this formalism includes all necessary information for a theory. Hence, it seems to be useful in order to generalize the conventional mechanics. A necessity for generalization may be explained by introducing dissipation as a reason of non-local interaction among particles \cite{schrodinger1914dynamik}, for its English translation, see \cite{065}. Such effects may be modeled by generalized continua \cite{eugster2017exegesis,eugster2018exegesisI,eugster2018exegesisII,dell2015origins}, especially at smaller length-scales, where the continuum length-scale converges to the microstructure \cite{075}. By using a variational method, generalized mechanics is acquired in a straight-forward manner \cite{gao2007variational,036}. However, its generalization to damage mechanics has difficulties, since damage mechanics is not acquired directly from the variational formalism. One possible approach is a hemivariational approach \cite{timofeev2021hemivariational,placidi2019simulation,placidi2018energy} but its extension in multiphysics is challenging. 

With this work, we expect to shed some light on this formalism and motivate to develop numerical methods based on a purely variational formulation \cite{074}. In this manner, possible explanations arise for difficult concepts such as contact formulations \cite{hanna2018partial}. We understand that the additional equations  from the extended \textsc{Noether} formalism is necessary to fulfill in order obtain physically correct models in fracture mechanics \cite{ruter2007duality,bird2019accurate}. These configurational forces are employed to compute the crack propagation in linear elasticity \cite{miehe2007computational} and elasto-plasticity \cite{ozencc2014evaluation}. Herein, we have derived the analogous equations for the generalized mechanics.

\section{Appendix} \label{appendix}

\subsection*{Transformation examples}
Suppose that $x^i=(x,y,z)$ refer to physical coordinates expressed in a Cartesian system and the transformation is the orthogonal rotation around $+z$-axis,
\begeq
x'=x\cos(\eps)+y\sin(\eps) \comma y'=-x\sin(\eps)+y\cos(\eps) \comma z'=z \ ,
\eqend
where $\eps$ is the rotation angle. Suppose that $x^i=(t,x,y,z)$ refer to space-time where the transformation is to a moving system in the direction of $x$. Between inertial systems, the following is called a \textsc{Galilei}an transformation:
\begeq
t'=t \comma x'=x-\eps t \comma y'=y \comma z'=z \comma
\eqend
where $\eps$ is the constant velocity. Suppose now that $x^i=(x,y,z,ct)$ is space-time in a \textsc{Minkowski}an system and a possible special \textsc{Lorentz} transformation reads
\begeq
x'=x \cosh(\eps) - ct \sinh(\eps) \comma y'=y \comma z'=z \comma ct'= -x\sinh(\eps) + ct\cosh(\eps) \ ,
\eqend
where $\eps$ is called rapidity of the transformation and the speed of light, $c$, is a universal constant, thus we have used $(ct)'=ct'$.

\subsection*{Measure's role in irreversibility}

We demonstrate in a simplified form how the measure gets a role in the irreversibility. Although the given example below is out of our scope in mechanics, it is beneficial to see this relation. Probably, this application is the only physical example, where the metric evolution is known. In cosmology  \cite{weinberg1972gravitation, misner1973gravitation, Susskind2} the universe is expanding with a (known) parameter $a=\bar a(t)$ everywhere the same, leading to the metric for that expanding universe:
\begeq
g_{ij}=\begin{pmatrix} 
a^2 & 0 & 0 \\ 
0 & a^2 & 0 \\ 
0 & 0 & a^2 
\end{pmatrix} 
\comma g=\det (g_{ij})=a^6 \comma \sqrt{g}=a^3 \ .
\eqend
Thus, the infinitesimal volume element reads $\d V = \sqrt g \d x = a^3 \d x$ and is time dependent. The rigid motion of galaxies, for example in one direction, $\chi$, will be calculated. We use the same short notation, $\Dt \chi = \p\chi/\p t$, and build the \textsc{Lagrange}an in that time dependent (expanding) metric
\begeq
L = \int \L \d x = \int \bigg( \frac12 \rho \Dt \chi \Dt \chi  - \mathcal V  \Big) a^3 \d x \ .
\eqend
The latter gives the energy density with the ground state $\mathcal V=\rho U \chi$ depending solely on the motion $\chi$, the ground state has different names in the literature: dark energy, vacuum energy, as well as cosmological constant. We plug in the \textsc{Lagrange} density into the \textsc{Euler--Lagrange} equations, use the material frame, $\Dt \rho=0$, and obtain
\begeq
\pd{\L}{\chi} - \pd{}{t}\pd{\L}{\Dt \chi} =0 \ ,\\
- \rho U a^3 - \pd{}{t} \Big( \rho \Dt \chi a^3 \Big) = 0 \ , \\ 
U a^3 + \DDt\chi a^3 + 3 a^2 \Dt a \Dt\chi  =0\\
U + \DDt\chi + 3 h \Dt\chi  = 0 \ ,
\eqend
where we have used the so-called \textsc{Hubble} constant $h=\Dt a/a$. Obviously, due to the expansion with velocity related to $h$, there is a damping in this partial differential equation such that the process is irreversible. We use this analogy and understand plasticity in mechanics as an irreversible change of the reference frame (herein the metric). Of course, the situation is far more difficult since additional governing equations need to be solved. In this simple example from cosmology, the \textsc{Hubble} constant is a given constant.

\subsection*{Acknowledgments}
B. Emek Abali had the pleasure of discussing the formalism with Reinhold Kienzler in an early stage of this work.

\bibliographystyle{special}
\bibliography{References}

\begin{thebibliography}{100}

\bibitem{truesdell_toupin}
C.~Truesdell \& R.~A. Toupin (1960) {\it Encyclopedia of physics, volume
  {III}/1, principles of classical mechanics and field theory\/}, chap. The
  classical field theories, pp. 226--790, Springer-Verlag,
  Berlin/G\"ottingen/Heidelberg

\bibitem{eckart1940I}
C.~Eckart (1940) {\it The thermodynamics of irreversible processes. i. the
  simple fluid\/}, Phys. Rev., {\bf 58}:pp. 267--269

\bibitem{eckart1940II}
C.~Eckart (1940) {\it The thermodynamics of irreversible processes. ii. fluid
  mixtures\/}, Physical Review, {\bf 58(3)}:p. 269

\bibitem{eckart1940III}
C.~Eckart (1940) {\it The thermodynamics of irreversible processes. iii.
  relativistic theory of the simple fluid\/}, Physical Review, {\bf 58(10)}:p.
  919

\bibitem{ColemanNoll1963}
B.~Coleman \& W.~Noll (1963) {\it The thermodynamics of elastic materials with
  heat conduction and viscosity\/}, Archive for Rational Mechanics and
  Analysis, {\bf 13(1)}:pp. 167--178

\bibitem{mueller1973}
I.~M{\"u}ller (1973) {\it Thermodynamik\/}, Bertelsmann-Universit{\"a}tsverlag

\bibitem{groot1984}
S.~R. de~Groot \& P.~Mazur (1984) {\it Non-equilibrium thermodynamics\/}, Dover
  Publications (New York)

\bibitem{muller1993extended}
I.~M{\"u}ller \& T.~Ruggeri (1993) {\it Extended thermodynamics\/}, Springer

\bibitem{muller2023electrodynamics}
I.~M{\"u}ller \& W.~H. M{\"u}ller (2023) {\it Electrodynamics and rational
  thermodynamics\/}, ZAMM-Journal of Applied Mathematics and
  Mechanics/Zeitschrift f{\"u}r Angewandte Mathematik und Mechanik, p.
  e202300209

\bibitem{014}
B.~E. Abali (2014) {\it Thermodynamically Compatible Modeling, Determination of
  Material Parameters, and Numerical Analysis of Nonlinear Rheological
  Materials\/}, Doctoral Thesis, Technische Universit\"at Berlin, epubli

\bibitem{dell2014complete}
F.~dell'Isola, G.~Maier, U.~Perego, U.~Andreaus, R.~Esposito, \& S.~Forest
  (2014) {\it The complete works of Gabrio Piola: Volume I: Commented English
  Translation\/}, vol.~38, Springer

\bibitem{dell2018complete}
F.~dell'Isola, U.~Andreaus, A.~Cazzani, R.~Esposito, L.~Placidi, U.~Perego,
  G.~Maier, \& P.~Seppecher (2018) {\it The Complete Works of Gabrio Piola:
  Volume II: Commented English Translation\/}, vol.~97, Springer

\bibitem{seliger1968variational}
R.~L. Seliger \& G.~B. Whitham (1968) {\it Variational principles in continuum
  mechanics\/}, Proceedings of the Royal Society of London. Series A.
  Mathematical and Physical Sciences, {\bf 305(1480)}:pp. 1--25

\bibitem{auffray2015analytical}
N.~Auffray, F.~dell’Isola, V.~A. Eremeyev, A.~Madeo, \& G.~Rosi (2015) {\it
  Analytical continuum mechanics {\`a} la {Hamilton--Piola} least action
  principle for second gradient continua and capillary fluids\/}, Mathematics
  and Mechanics of Solids, {\bf 20(4)}:pp. 375--417

\bibitem{kock1991fluid}
E.~Kock \& L.~Olson (1991) {\it Fluid-structure interaction analysis by the
  finite element method--a variational approach\/}, International journal for
  numerical methods in engineering, {\bf 31(3)}:pp. 463--491

\bibitem{placidi2008variational}
L.~Placidi, F.~dell'Isola, N.~Ianiro, \& G.~Sciarra (2008) {\it Variational
  formulation of pre-stressed solid--fluid mixture theory, with an application
  to wave phenomena\/}, European Journal of Mechanics-A/Solids, {\bf 27(4)}:pp.
  582--606

\bibitem{prix2004variational}
R.~Prix (2004) {\it Variational description of multifluid hydrodynamics:
  Uncharged fluids\/}, Physical Review D, {\bf 69(4)}:p. 043001

\bibitem{bekenstein2000conservation}
J.~D. Bekenstein \& A.~Oron (2000) {\it Conservation of circulation in
  magnetohydrodynamics\/}, Physical Review E, {\bf 62(4)}:p. 5594

\bibitem{altenbach2014vibration}
H.~Altenbach \& V.~A. Eremeyev (2014) {\it Vibration analysis of non-linear
  6-parameter prestressed shells\/}, Meccanica, {\bf 49(8)}:pp. 1751--1761

\bibitem{eremeyev2019dynamic}
V.~A. Eremeyev (2019) {\it On dynamic boundary conditions within the linear
  {Steigmann-Ogden} model of surface elasticity and strain gradient
  elasticity\/}, H.~Altenbach, A.~Belyaev, V.~Eremeyev, A.~Krivtsov, \&
  A.~Porubov (Eds.) {\it Dynamical Processes in Generalized Continua and
  Structures\/}, vol. 103 of {\it Advanced Structured Materials\/}, pp.
  195--207, Springer, Cham

\bibitem{placidi2015variational}
L.~Placidi (2015) {\it A variational approach for a nonlinear 1-dimensional
  second gradient continuum damage model\/}, Continuum Mechanics and
  Thermodynamics, {\bf 27(4)}:pp. 623--638

\bibitem{ciallella2021rate}
A.~Ciallella, D.~Pasquali, M.~Go{\l}aszewski, F.~D’Annibale, \& I.~Giorgio
  (2021) {\it A rate-independent internal friction to describe the hysteretic
  behavior of pantographic structures under cyclic loads\/}, Mechanics Research
  Communications, {\bf 116}:p. 103761

\bibitem{steigmann1996variational}
D.~J. Steigmann (1996) {\it The variational structure of a nonlinear theory for
  spatial lattices\/}, Meccanica, {\bf 31(4)}:pp. 441--455

\bibitem{bourdin2008variational}
B.~Bourdin, G.~A. Francfort, \& J.-J. Marigo (2008) {\it The variational
  approach to fracture\/}, Journal of elasticity, {\bf 91(1-3)}:pp. 5--148

\bibitem{francfort1998revisiting}
G.~A. Francfort \& J.-J. Marigo (1998) {\it Revisiting brittle fracture as an
  energy minimization problem\/}, Journal of the Mechanics and Physics of
  Solids, {\bf 46(8)}:pp. 1319--1342

\bibitem{vujanovic1975variational}
B.~Vujanovic (1975) {\it A variational principle for non-conservative dynamical
  systems\/}, ZAMM-Journal of Applied Mathematics and Mechanics/Zeitschrift
  Angewandte Mathematik und Mechanik, {\bf 55}:pp. 321--331

\bibitem{lanczos1949variational}
C.~Lanczos (1949) {\it The Variational Principles of Mechanics\/}, University
  of Toronto Press

\bibitem{gelfand1963calculus}
I.~Gelfand \& S.~Fomin (1963) {\it Calculus of Variations\/}, Prentice-Hall
  Inc., Englewood Cliffs

\bibitem{goldstein2002classical}
H.~Goldstein, C.~Poole, \& J.~Safko (2002) {\it Classical mechanics\/},
  American Association of Physics Teachers

\bibitem{arnold2021topological}
V.~I. Arnold \& B.~A. Khesin (2021) {\it Topological methods in
  hydrodynamics\/}, vol. 125 of {\it Applied Mathematical Sciences\/}, Springer
  Nature

\bibitem{marsden2001variational}
J.~E. Marsden, S.~Pekarsky, S.~Shkoller, \& M.~West (2001) {\it Variational
  methods, multisymplectic geometry and continuum mechanics\/}, Journal of
  Geometry and Physics, {\bf 38(3-4)}:pp. 253--284

\bibitem{morrison1984bracket}
P.~J. Morrison (1984) {\it Bracket formulation for irreversible classical
  fields\/}, Physics Letters A, {\bf 100(8)}:pp. 423--427

\bibitem{grmela1986bracket}
M.~Grmela (1986) {\it Bracket formulation of diffusion-convection equations\/},
  Physica D: Nonlinear Phenomena, {\bf 21(2-3)}:pp. 179--212

\bibitem{holm1998euler}
D.~D. Holm, J.~E. Marsden, \& T.~S. Ratiu (1998) {\it The {Euler--P}oincar{\'e}
  equations and semidirect products with applications to continuum theories\/},
  Advances in Mathematics, {\bf 137(1)}:pp. 1--81

\bibitem{gay2018lagrangian}
F.~Gay-Balmaz \& H.~Yoshimura (2018) {\it From {L}agrangian mechanics to
  nonequilibrium thermodynamics: a variational perspective\/}, Entropy, {\bf
  21(1)}:p.~8

\bibitem{beris1994thermodynamics}
A.~N. Beris \& B.~J. Edwards (1994) {\it Thermodynamics of flowing systems:
  with internal microstructure\/}, no.~36 in Oxford Engineering Science Series,
  Oxford University Press

\bibitem{pavelka2018multiscale}
M.~Pavelka, V.~Klika, \& M.~Grmela (2018) {\it Multiscale thermo-dynamics\/},
  {\it Multiscale Thermo-Dynamics\/}, de Gruyter

\bibitem{eshelby1951force}
J.~D. Eshelby (1951) {\it The force on an elastic singularity\/}, Phil. Trans.
  R. Soc. Lond. A, {\bf 244(877)}:pp. 87--112

\bibitem{eshelby1975elastic}
J.~D. Eshelby (1975) {\it The elastic energy-momentum tensor\/}, Journal of
  elasticity, {\bf 5(3-4)}:pp. 321--335

\bibitem{dirac}
P.~A.~M. Dirac (1975) {\it General theory of relativity\/}, Princeton
  University Press (1996)

\bibitem{rice1968path}
J.~R. Rice (1968) {\it A path independent integral and the approximate analysis
  of strain concentration by notches and cracks\/}, Journal of applied
  mechanics, {\bf 35(2)}:pp. 379--386

\bibitem{altenbach1990konzepte}
H.~Altenbach, J.~Altenbach, \& P.~Schie{\ss{}}e (1990) {\it Konzepte der
  {S}ch{\"a}digungsmechanik und ihre {A}nwendung bei der werkstoffmechanischen
  {B}auteilanalyse\/}, Technische Mechanik, {\bf 11(2)}:pp. 81--93

\bibitem{kroner1993configurational}
E.~Kr{\"o}ner (1993) {\it Configurational and material forces in the theory of
  defects in ordered structures\/}, {\it Materials Science Forum\/}, vol. 123,
  pp. 447--454

\bibitem{maugin1995material}
G.~A. Maugin (1995) {\it Material forces: concepts and applications\/}, Applied
  Mechanics Reviews, {\bf 48(5)}:pp. 213--245

\bibitem{gurtin1996configurational}
M.~E. Gurtin \& P.~Podio-Guidugli (1996) {\it Configurational forces and the
  basic laws for crack propagation\/}, Journal of the Mechanics and Physics of
  Solids, {\bf 44(6)}:pp. 905--927

\bibitem{steinmann2000application}
P.~Steinmann (2000) {\it Application of material forces to hyperelastostatic
  fracture mechanics. i. continuum mechanical setting\/}, International Journal
  of Solids and Structures, {\bf 37(48-50)}:pp. 7371--7391

\bibitem{kienzler2002fracture}
R.~Kienzler \& G.~Herrmann (2002) {\it Fracture criteria based on local
  properties of the eshelby tensor\/}, Mechanics Research Communications, {\bf
  29(6)}:pp. 521--527

\bibitem{li2006dual}
S.~Li \& A.~Gupta (2006) {\it On dual configurational forces\/}, Journal of
  Elasticity, {\bf 84(1)}:pp. 13--31

\bibitem{kuhn2010continuum}
C.~Kuhn \& R.~Mueller (2010) {\it A continuum phase field model for
  fracture\/}, Engineering Fracture Mechanics, {\bf 77(18)}:pp. 3625--3634

\bibitem{agiasofitou2017micromechanics}
E.~Agiasofitou \& M.~Lazar (2017) {\it Micromechanics of dislocations in
  solids: {J-, M-, and L}-integrals and their fundamental relations\/},
  International Journal of Engineering Science, {\bf 114}:pp. 16--40

\bibitem{perez2017fracture}
N.~Perez (2017) {\it Fracture mechanics\/}, Springer

\bibitem{maugin2016fracture}
G.~A. Maugin (2016) {\it {Fracture: To Crack or Not to Crack. That Is the
  Question}\/}, {\it Continuum Mechanics through the Ages-From the Renaissance
  to the Twentieth Century\/}, vol. 223 of {\it Solid Mechanics and Its
  Applications\/}, pp. 215--242, Springer, Cham

\bibitem{dell1987derivation}
F.~dell'Isola \& A.~Romano (1987) {\it On the derivation of thermomechanical
  balance equations for continuous systems with a nonmaterial interface\/},
  International Journal of Engineering Science, {\bf 25(11-12)}:pp. 1459--1468

\bibitem{kienzler2012mechanics}
R.~Kienzler \& G.~Herrmann (2012) {\it Mechanics in material space: with
  applications to defect and fracture mechanics\/}, Springer Science \&
  Business Media

\bibitem{buratti2003}
G.~Buratti, Y.~Huo, \& I.~M{\"u}ller (2003) {\it Eshelby tensor as a tensor of
  free enthalpy\/}, Journal of elasticity, {\bf 72(1-3)}:pp. 31--42

\bibitem{wolff2018application}
M.~Wolff, M.~B{\"o}hm, \& H.~Altenbach (2018) {\it Application of the
  {M{\"u}ller--Liu} entropy principle to gradient-damage models in the
  thermo-elastic case\/}, International Journal of Damage Mechanics, {\bf
  27(3)}:pp. 387--408

\bibitem{mueller2002material}
R.~Mueller \& G.~Maugin (2002) {\it On material forces and finite element
  discretizations\/}, Computational mechanics, {\bf 29(1)}:pp. 52--60

\bibitem{ruter2006goal}
M.~R{\"u}ter \& E.~Stein (2006) {\it Goal-oriented a posteriori error estimates
  in linear elastic fracture mechanics\/}, Computer methods in applied
  mechanics and engineering, {\bf 195(4-6)}:pp. 251--278

\bibitem{singh2021pseudomomentum}
H.~Singh \& J.~Hanna (2021) {\it Pseudomomentum: origins and consequences\/},
  Zeitschrift f{\"u}r angewandte Mathematik und Physik (ZAMP), {\bf 72(3)}:pp.
  1--25

\bibitem{eckart1960}
C.~Eckart (1960) {\it Variation principles in hydrodynamics\/}, The physics of
  fluids, {\bf 3}:pp. 421--427

\bibitem{disalle2009}
R.~DiSalle (2009) {\it Space and time: Inertial frames\/}, E.~N. Zalta (Ed.)
  {\it The Stanford Encyclopedia of Philosophy\/}, winter 2009 edn.

\bibitem{noether1}
E.~Noether (1918) {\it Invarianten beliebiger {D}ifferentialausdr\"ucke\/},
  Nachr. v. d. Ges. d. Wiss. zu Goettingen 1918, p. 37-44

\bibitem{noether2}
E.~Noether (1918) {\it Invariante {V}ariationsprobleme\/}, Nachr. v. d. Ges. d.
  Wiss. zu Goettingen 1918, p. 235-257

\bibitem{kosmann2011noether}
Y.~Kosmann-Schwarzbach (2011) {\it The Noether Theorems\/}, Springer

\bibitem{collins1899}
J.~V. Collins (1899) {\it An elementary exposition of {G}rassmann's
  ``{A}usdehnungslehre,'' or theory of extension\/}, The American Mathematical
  Monthly, {\bf 6(8/9)}:pp. pp. 193--198

\bibitem{flanders}
H.~Flanders (1963) {\it Differential forms with applications to the physical
  sciences\/}, Dover Publications, Inc., New York (1989)

\bibitem{pauli2}
W.~Pauli (1921) {\it Theory of relativity\/}, Dover (1981) repub. of Pergamon
  Press, Oxford (1958)

\bibitem{neuenschwander2010}
D.~Neuenschwander (2010) {\it Emmy {N}oether's wonderful theorem\/}, Johns
  Hopkins University Press

\bibitem{helmholtz1887}
H.~von Helmholtz (1887) {\it {\"{U}}ber die physikalische {B}edeutung des
  {P}rincips der kleinsten {W}irkung\/}, Journal für die reine und angewandte
  Mathematik (Crelle's Journal), {\bf 100}:pp. 137--166

\bibitem{helmholtz1887ueber}
H.~von Helmholtz (1887) {\it {\"{U}}ber die physikalische {B}edeutung des
  {P}rincips der kleinsten {W}irkung (fortsetzung)\/}, Journal für die reine
  und angewandte Mathematik (Crelle's Journal), {\bf 100}:pp. 213--222

\bibitem{laue1911}
M.~Laue (1911) {\it {Zur Dynamik der R}elativit{\"a}tstheorie\/}, Annalen der
  Physik, {\bf 340(8)}:pp. 524--542

\bibitem{bessel1921}
E.~Bessel-Hagen (1921) {\it {\"U}ber die {E}rhaltungss{\"a}tze der
  {E}lektrodynamik\/}, Mathematische Annalen, {\bf 84(3-4)}:pp. 258--276

\bibitem{rund1966hamilton}
H.~Rund (1966) {\it The {Hamilton--Jacobi} theory in the calculus of
  variations: its role in mathematics and physics\/}, Krieger Publishing
  Company

\bibitem{trautman1967noether}
A.~Trautman (1967) {\it Noether equations and conservation laws\/},
  Communications in mathematical Physics, {\bf 6(4)}:pp. 248--261

\bibitem{rund1972direct}
H.~Rund (1972) {\it A direct approach to {N}oether’s theorem in the calculus
  of variations\/}, Utilitas Math, {\bf 2}:pp. 205--214

\bibitem{kobe2013noether}
D.~H. Kobe (2013) {\it Noether's theorem and the work-energy theorem for a
  charged particle in an electromagnetic field\/}, American Journal of Physics,
  {\bf 81(3)}:pp. 186--189

\bibitem{marsden2013introduction}
J.~E. Marsden \& T.~S. Ratiu (2013) {\it Introduction to mechanics and
  symmetry: a basic exposition of classical mechanical systems\/}, vol.~17,
  Springer Science \& Business Media

\bibitem{bersani2020lagrangian}
A.~M. Bersani \& P.~Caressa {\it Lagrangian descriptions of dissipative
  systems: a review\/}, Mathematics and Mechanics of Solids, p.
  1081286520971834

\bibitem{eshelby1956continuum}
J.~Eshelby (1956) {\it The continuum theory of lattice defects\/}, {\it Solid
  state physics\/}, vol.~3, pp. 79--144, Elsevier

\bibitem{knowles1973asymptotic}
J.~K. Knowles \& E.~Sternberg (1973) {\it An asymptotic finite-deformation
  analysis of the elastostatic field near the tip of a crack\/}, Journal of
  Elasticity, {\bf 3(2)}:pp. 67--107

\bibitem{maugin2020material}
G.~A. Maugin (1993) {\it Material inhomogeneities in elasticity\/}, Chapman \&
  Hall, London

\bibitem{gurses2009computational}
E.~G{\"u}rses \& C.~Miehe (2009) {\it A computational framework of
  three-dimensional configurational-force-driven brittle crack propagation\/},
  Computer Methods in Applied Mechanics and Engineering, {\bf 198(15-16)}:pp.
  1413--1428

\bibitem{fletcher1976conservation}
D.~C. Fletcher (1976) {\it Conservation laws in linear elastodynamics\/},
  Archive for Rational Mechanics and Analysis, {\bf 60(4)}:pp. 329--353

\bibitem{kalpakides2004canonical}
V.~Kalpakides \& G.~Maugin (2004) {\it Canonical formulation and conservation
  laws of thermoelasticity without dissipation\/}, Reports on Mathematical
  Physics, {\bf 53(3)}:pp. 371--391

\bibitem{049}
F.~dell'Isola, P.~Seppecher, M.~Spagnuolo, E.~Barchiesi, F.~Hild, T.~Lekszycki,
  I.~Giorgio, L.~Placidi, U.~Andreaus, M.~Cuomo, S.~R. Eugster, A.~Pfaff,
  K.~Hoschke, R.~Langkemper, E.~Turco, R.~Sarikaya, A.~Misra, M.~De~Angelo,
  F.~D'Annibale, A.~Bouterf, X.~Pinelli, A.~Misra, B.~Desmorat, M.~Pawlikowski,
  C.~Dupuy, D.~Scerrato, P.~Peyre, M.~Laudato, L.~Manzari, P.~Göransson,
  C.~Hesch, S.~Hesch, P.~Franciosi, J.~Dirrenberger, F.~Maurin, Z.~Vangelatos,
  C.~Grigoropoulos, V.~Melissinaki, M.~Farsari, W.~Muller, B.~E. Abali,
  C.~Liebold, G.~Ganzosch, P.~Harrison, R.~Drobnicki, L.~Igumnov, F.~Alzahrani,
  \& T.~Hayat (2019) {\it Advances in pantographic structures: design,
  manufacturing, models, experiments and image analyses\/}, Continuum Mechanics
  and Thermodynamics, {\bf 31(4)}:pp. 1231--1282

\bibitem{muller2020experimental}
W.~H. M{\"u}ller (2020) {\it The experimental evidence for higher gradient
  theories\/}, {\it Mechanics of Strain Gradient Materials\/}, pp. 1--18,
  Springer

\bibitem{askes2011gradient}
H.~Askes \& E.~C. Aifantis (2011) {\it Gradient elasticity in statics and
  dynamics: an overview of formulations, length scale identification
  procedures, finite element implementations and new results\/}, International
  Journal of Solids and Structures, {\bf 48(13)}:pp. 1962--1990

\bibitem{auffray2010strain}
N.~Auffray, R.~Bouchet, \& Y.~Brechet (2010) {\it Strain gradient elastic
  homogenization of bidimensional cellular media\/}, International Journal of
  Solids and Structures, {\bf 47(13)}:pp. 1698--1710

\bibitem{tran2012micromechanics}
T.-H. Tran, V.~Monchiet, \& G.~Bonnet (2012) {\it A micromechanics-based
  approach for the derivation of constitutive elastic coefficients of
  strain-gradient media\/}, International Journal of Solids and Structures,
  {\bf 49(5)}:pp. 783--792

\bibitem{barboura2018establishment}
S.~Barboura \& J.~Li (2018) {\it Establishment of strain gradient constitutive
  relations by using asymptotic analysis and the finite element method for
  complex periodic microstructures\/}, International Journal of Solids and
  Structures, {\bf 136}:pp. 60--76

\bibitem{057}
B.~E. Abali, H.~Yang, \& P.~Papadopoulos (2019) {\it A computational approach
  for determination of parameters in generalized mechanics\/}, H.~Altenbach,
  W.~H. M\"uller, \& B.~E. Abali (Eds.) {\it Higher Gradient Materials and
  Related Generalized Continua\/}, Advanced Structured Materials, vol. 120,
  chap.~1, pp. 1--18, Springer, Cham

\bibitem{069}
B.~E. Abali \& E.~Barchiesi (2021) {\it Additive manufacturing introduced
  substructure and computational determination of metamaterials parameters by
  means of the asymptotic homogenization\/}, Continuum Mechanics and
  Thermodynamics, {\bf 33}:pp. 993--1009

\bibitem{solyaev2022self}
Y.~Solyaev (2022) {\it Self-consistent assessments for the effective properties
  of two-phase composites within strain gradient elasticity\/}, Mechanics of
  Materials, {\bf 169}:p. 104321

\bibitem{084}
B.~Vazic, B.~E. Abali, H.~Yang, \& P.~Newell (2022) {\it Mechanical analysis of
  heterogeneous materials with higher-order parameters\/}, Engineering with
  Computers, {\bf 38(6)}:pp. 5051--5067

\bibitem{085}
H.~Yang, B.~E. Abali, W.~H. Müller, S.~Barboura, \& J.~Li (2022) {\it
  Verification of asymptotic homogenization method developed for periodic
  architected materials in strain gradient continuum\/}, International Journal
  of Solids and Structures, {\bf 238}:p. 111386

\bibitem{yvonnet2020computational}
J.~Yvonnet, N.~Auffray, \& V.~Monchiet (2020) {\it Computational second-order
  homogenization of materials with effective anisotropic strain-gradient
  behavior\/}, International Journal of Solids and Structures, {\bf 191}:pp.
  434--448

\bibitem{lahbazi2022size}
A.~Lahbazi, I.~Goda, \& J.-F. Ganghoffer (2022) {\it Size-independent strain
  gradient effective models based on homogenization methods: Applications to 3d
  composite materials, pantograph and thin walled lattices\/}, Composite
  Structures, {\bf 284}:p. 115065

\bibitem{areias2022finite}
P.~Areias, R.~Melicio, F.~Carapau, \& J.~Carrilho~Lopes (2022) {\it Finite
  gradient models with enriched rbf-based interpolation\/}, Mathematics, {\bf
  10(16)}:p. 2876

\bibitem{mindlin1968first}
R.~D. Mindlin \& N.~Eshel (1968) {\it On first strain-gradient theories in
  linear elasticity\/}, International Journal of Solids and Structures, {\bf
  4(1)}:pp. 109--124

\bibitem{altan1997some}
B.~Altan \& E.~Aifantis (1997) {\it On some aspects in the special theory of
  gradient elasticity\/}, Journal of the Mechanical Behavior of Materials, {\bf
  8(3)}:pp. 231--282

\bibitem{082}
N.~Shekarchizadeh, M.~Laudato, L.~Manzari, B.~E. Abali, I.~Giorgio, \& A.~M.
  Bersani (2021) {\it Parameter identification of a second-gradient model for
  the description of pantographic structures in dynamic regime\/}, Zeitschrift
  f{\"u}r angewandte Mathematik und Physik (ZAMP), {\bf 72(6)}:p. 190

\bibitem{schrodinger1914dynamik}
E.~Schr{\"o}dinger (1914) {\it Zur {D}ynamik elastisch gekoppelter
  {P}unktsysteme\/}, Annalen der Physik, {\bf 349(14)}:pp. 916--934

\bibitem{065}
U.~M{\"u}hlich, B.~E. Abali, \& F.~dell'Isola (2020) {\it Commented translation
  of {E}rwin {S}chr{\"o}dinger's paper '{O}n the dynamics of elastically
  coupled point systems'({Z}ur {D}ynamik elastisch gekoppelter
  {P}unktsysteme)\/}, Mathematics and Mechanics of Solids, {\bf 26(1)}:p.
  1081286520942955

\bibitem{eugster2017exegesis}
S.~R. Eugster \& F.~dell'Isola (2017) {\it Exegesis of the introduction and
  sect. i from ``{Fundamentals of the Mechanics of Continua}'' by {E.
  H}ellinger\/}, ZAMM-Journal of Applied Mathematics and Mechanics/Zeitschrift
  f{\"u}r Angewandte Mathematik und Mechanik, {\bf 97(4)}:pp. 477--506

\bibitem{eugster2018exegesisI}
S.~R. Eugster \& F.~dell'Isola (2018) {\it Exegesis of sect. ii and iii. a from
  “{Fundamentals of the Mechanics of Continua}” by {E. H}ellinger\/},
  ZAMM-Journal of Applied Mathematics and Mechanics/Zeitschrift f{\"u}r
  Angewandte Mathematik und Mechanik, {\bf 98(1)}:pp. 31--68

\bibitem{eugster2018exegesisII}
S.~R. Eugster \& F.~dell'Isola (2018) {\it Exegesis of sect. iii. b from
  “{Fundamentals of the Mechanics of Continua}” by {E. H}ellinger\/},
  ZAMM-Journal of Applied Mathematics and Mechanics/Zeitschrift f{\"u}r
  Angewandte Mathematik und Mechanik, {\bf 98(1)}:pp. 69--105

\bibitem{dell2015origins}
F.~dell'Isola, U.~Andreaus, \& L.~Placidi (2015) {\it At the origins and in the
  vanguard of peridynamics, non-local and higher-gradient continuum mechanics:
  an underestimated and still topical contribution of {Gabrio Piola}\/},
  Mathematics and Mechanics of Solids, {\bf 20(8)}:pp. 887--928

\bibitem{075}
K.~K. Mandadapu, B.~E. Abali, \& P.~Papadopoulos (2021) {\it On the polar
  nature and invariance properties of a thermomechanical theory for
  continuum-on-continuum homogenization\/}, Mathematics and Mechanics of
  Solids, {\bf 26(11)}:pp. 1581--1598

\bibitem{gao2007variational}
X.-L. Gao \& S.~Park (2007) {\it Variational formulation of a simplified strain
  gradient elasticity theory and its application to a pressurized thick-walled
  cylinder problem\/}, International Journal of Solids and Structures, {\bf
  44(22-23)}:pp. 7486--7499

\bibitem{036}
B.~E. Abali (2018) {\it Revealing the physical insight of a length scale
  parameter in metamaterials by exploring the variational formulation\/},
  Continuum Mechanics and Thermodynamics, {\bf 31(4)}:pp. 885--894

\bibitem{timofeev2021hemivariational}
D.~Timofeev, E.~Barchiesi, A.~Misra, \& L.~Placidi (2021) {\it Hemivariational
  continuum approach for granular solids with damage-induced anisotropy
  evolution\/}, Mathematics and Mechanics of Solids, {\bf 26(5)}:pp. 738--770

\bibitem{placidi2019simulation}
L.~Placidi, A.~Misra, \& E.~Barchiesi (2019) {\it Simulation results for damage
  with evolving microstructure and growing strain gradient moduli\/}, Continuum
  Mechanics and Thermodynamics, {\bf 31(4)}:pp. 1143--1163

\bibitem{placidi2018energy}
L.~Placidi \& E.~Barchiesi (2018) {\it Energy approach to brittle fracture in
  strain-gradient modelling\/}, Proceedings of the Royal Society A:
  Mathematical, Physical and Engineering Sciences, {\bf 474(2210)}:p. 20170878

\bibitem{074}
B.~E. Abali, A.~Klunker, E.~Barchiesi, \& L.~Placidi (2021) {\it A novel
  phase-field approach to brittle damage mechanics of gradient metamaterials
  combining action formalism and history variable\/}, ZAMM-Journal of Applied
  Mathematics and Mechanics/Zeitschrift f{\"u}r Angewandte Mathematik und
  Mechanik, {\bf 101(9)}:p. e202000289

\bibitem{hanna2018partial}
J.~Hanna, H.~Singh, \& E.~Virga (2018) {\it Partial constraint singularities in
  elastic rods\/}, Journal of Elasticity, {\bf 133(1)}:pp. 105--118

\bibitem{ruter2007duality}
M.~R{\"u}ter \& E.~Stein (2007) {\it On the duality of finite element
  discretization error control in computational newtonian and eshelbian
  mechanics\/}, Computational Mechanics, {\bf 39}:pp. 609--630

\bibitem{bird2019accurate}
R.~Bird, W.~M. Coombs, \& S.~Giani (2019) {\it Accurate configuration force
  evaluation via hp-adaptive discontinuous galerkin finite element analysis\/},
  Engineering Fracture Mechanics, {\bf 216}:p. 106370

\bibitem{miehe2007computational}
C.~Miehe, E.~G{\"u}rses, \& M.~Birkle (2007) {\it A computational framework of
  configurational-force-driven brittle fracture based on incremental energy
  minimization\/}, International Journal of Fracture, {\bf 145}:pp. 245--259

\bibitem{ozencc2014evaluation}
K.~{\"O}zen{\c{c}}, M.~Kaliske, G.~Lin, \& G.~Bhashyam (2014) {\it Evaluation
  of energy contributions in elasto-plastic fracture: a review of the
  configurational force approach\/}, Engineering Fracture Mechanics, {\bf
  115}:pp. 137--153

\bibitem{weinberg1972gravitation}
S.~Weinberg (1972) {\it Gravitation and cosmology: principles and applications
  of the general theory of relativity\/}, Wiley, New York

\bibitem{misner1973gravitation}
C.~W. Misner, K.~S. Thorne, \& J.~A. Wheeler (1973) {\it Gravitation\/},
  Macmillan

\bibitem{Susskind2}
L.~Susskind (2009), {\it Cosmology\/}, video lectures in Stanford,
  academicearth.org

\end{thebibliography}

\end{document}